\documentclass[aps,prbbib,twocolumn,epsf]{revtex4}

\usepackage{graphicx}
\usepackage[usenames]{color}

\begin{document}
\draft

\title{Optical spin transfer and spin-orbit torques in thin film ferromagnets}
\author{Junwen Li$^{1,2}$ and Paul M. Haney$^1$}

\affiliation{1.  Center for Nanoscale Science and Technology, National Institute of Standards and Technology, Gaithersburg, MD 20899 \\
2.  Maryland NanoCenter, University of Maryland, College Park, MD 20742, USA
}
\begin{abstract}
We study the optically induced torques in thin film ferromagnetic layers under excitation by circularly polarized light.   We study cases both with and without Rashba spin-orbit coupling using a 4-band model.  In the absence of Rashba spin-orbit coupling, we derive an analytic expression for the optical torques, revealing the conditions under which the torque is mostly derived from optical spin transfer torque (i.e. when the torque is along the direction of optical angular momentum), versus when the torque is derived from the inverse Faraday effect (i.e. when the torque is perpendicular to the optical angular momentum).  We find the optical spin transfer torque dominates provided that the excitation energy is far away from band edge transitions, and the magnetic exchange splitting is much greater than the lifetime broadening.  For the case with large Rashba spin-orbit coupling and out-of-plane magnetization, we find the torque is generally perpendicular to the photon angular momentum and is ascribed to an optical Edelstein effect.
\end{abstract}

\maketitle

\section{Introduction}

The interaction between light and magnetism is of fundamental and technological interest \cite{kirilyuk2010ultrafast}.  There are several mechanisms underlying optical control of magnetism.  Among these include a range of thermal and quantum mechanical effects which lead to ultrafast demagnetization \cite{zhang2000laser,lambert2014all,guidoni2002magneto,battiato2010superdiffusive}.  Light absorption also modifies the electron distribution function, which can change the magnetic anisotropy and lead to magnetic dynamics \cite{duong2004ultrafast,hashimoto2008photoinduced,tesavrova2013experimental}.  Another optical effect is spin transfer torque from absorption of circularly polarized light.  Optical spin transfer torque operates on the same principle as electrical current-induced spin transfer torque \cite{fernandez2003optical,chovan2006ultrafast,nvemec2012experimental}: In both cases conservation of total spin angular momentum implies that a net flux of angular momentum flow into a ferromagnet results in a torque on the magnetization \cite{slonczewski1996current,berger1996emission,ralph2008spin}.  In this simple picture of optical spin transfer torque, spin-orbit coupling is neglected so that total spin is conserved. In instances where spin-orbit coupling is not negligible, spin conservation no longer applies and an excitation may induce a torque on the magnetization in which the angular momentum is supplied by the lattice \cite{haney2010current,qaiumzadeh2013manipulation}.  The angular momentum transfer is mediated by spin-orbit coupling, and the resulting torque is known as spin-orbit torque.  Spin-orbit torques have been realized and extensively studied using DC electrical excitation of heavy metal-ferromagnet bilayers \cite{miron2011perpendicular,garello2013symmetry,kim2013layer}, and in individual Rahsba ferromagnets \cite{qaiumzadeh2015spin}.  Spin-orbit torque may have advantages over spin transfer torque in terms of the efficiency of magnetic switching \cite{liu2012spin}.  Optically excited spin-orbit torques are so far less well established.

The most well studied system for optical magnetic torques is GaMnAs, an archetypical ferromagnetic semiconductor.  To our knowledge, all previous analysis of GaMnAs consider the bulk response \cite{kapetanakis2009femtosecond,fernandez2003optical,chovan2006ultrafast,qaiumzadeh2013manipulation}.  The optical response of metallic systems has also recently been studied theoretically \cite{berritta2016ab,freimuth2016laser}.
These previously considered semiconductor and metallic bulk systems naturally lack Rashba spin-orbit coupling.  Rashba spin-orbit coupling is central to the analysis of Ref. \cite{qaiumzadeh2016theory}, which computes the optically induced effective magnetic fields in a thin film Rashba ferromagnet with parallel $\bf L$ and $\bf M$.

\begin{figure}[!h]
  \includegraphics[scale=0.23]{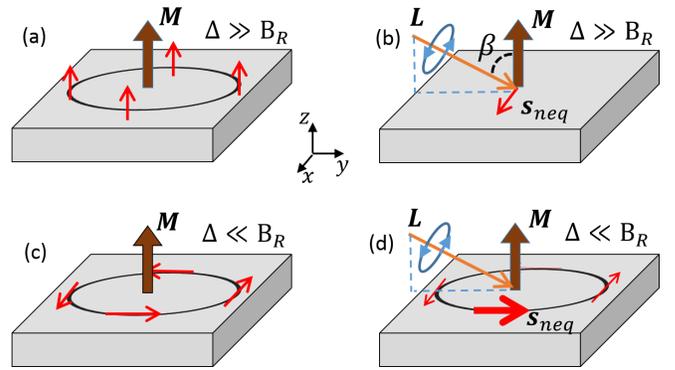}
  \caption{(a) shows magnetization (solid dark arrow) and spin of eigenstates (thin red arrows) for magnetic exchange splitting $\Delta$ much greater than Rashba spin-orbit splitting $B_{\rm R}$.  (b) shows the nonequilibrium spin density induced by light with circular polarization ${\bf L}$ incident on the sample at an angle $\beta$ with respect to the $z$-axis.   (c) and (d) show the same in the case $\Delta \ll B_{\rm R}$. }
    \label{fig:schematic}
\end{figure}

Here we consider the optical spin-transfer and spin-orbit torques which are present in a thin film ferromagnet, both with and without Rashba spin-orbit coupling.
The geometry is shown in Fig. \ref{fig:schematic}: we study both in-plane and out-of-plane magnetic configurations, and consider light with an oblique angle of incidence $\beta$.  Similar geometries have been employed in recent experiments \cite{gorchon2017single,choi2017optical}.  The misalignment between $\bf L$ and $\bf M$ is a key distinction between this work and Ref. \cite{qaiumzadeh2016theory}, and is responsible for the magnetic torques analyzed here.  The torque on the magnetization is perpendicular to ${\bf M}$ and can be written in terms of the vector components along ${\bf M}\times{\bf L}$ and ${\bf M}\times{\bf M}\times{\bf L}$.  These torques can be understood arising from effective magnetic fields: the torque in the ${\bf M}\times{\bf L}$ direction is the result of an effective magnetic field along ${\bf L}$.  This field is identified with the inverse Faraday effect \cite{pershan1966theoretical,kimel2005ultrafast}, and we note that the associated torque is odd with both helicity and magnetization.  The torque along the ${\bf M}\times{\bf M}\times{\bf L}$ is due to an effective field along the ${\bf M}\times{\bf L}$ direction.  This torque is identified as the optical spin transfer torque \cite{fernandez2003optical,freimuth2016laser}, and is odd in ${\bf L}$ and even in ${\bf M}$.

The relative magnitude of these two components of the torque plays an important role in interpreting experiments and determining the underlying mechanisms of the torque.  Ref. \cite{choi2017optical} measures the relative magnitude of the optical torques derived from the spin transfer and inverse Faraday effects for 10 nm thick layers of Ni, Co, and Fe.  They find the torque derived from the inverse Faraday effect is larger, although the addition of a Pt capping layer substantially increases the optical spin transfer torque component.  Ref. \cite{freimuth2016laser} uses density functional theory and a Keldysh Green's function approach to compute the components of the optical torque as a function of lifetime broadening for bulk Co, Fe, and FePt.  They find the torque associated with the inverse Faraday effect is generally larger than the optical spin transfer torque for Co and FePt, but that the two components are comparable for Fe.  There is good agreement between the magnitude between experiment and theory for the inverse Faraday effect-derived torque for Fe.  It's difficult to rationalize the relative magnitude of the torque in first principles calculations due to the complexity of the electronic band structure.

In this work we use time-dependent perturbation theory to derive expressions for the steady state torque.  This approach has the benefit of providing a basis for understanding the relative magnitude of the torque components in terms of single particle wave functions and energies.  We first present analytic expressions for the optical torques in the absence of Rashba spin-orbit coupling for a simple 4-band model.  The simplicity of the model enables closed form expressions, which provide insight into how system parameters determine the relative magnitudes of the optical spin transfer torque and the torque derived from the inverse Faraday effect.  We next compute the optical torque for strong Rashba spin-orbit coupling.  For an out-of-plane magnetization, the torque is primarily along the ${\bf M}\times{\bf L}$ direction.  In this case, the torque can be understood in similar terms as the Rashba-Edelstein derived torque \cite{manchon2008theory,garate2009influence}: Absorption of obliquely incident ($\beta\neq 0$) circularly polarized light with angular momentum ${\bf L}$ leads to an asymmetry in the electron distribution function in ${\bf k}$-space due to optical selection rules (see Fig. \ref{fig:bands}).  This asymmetric distribution leads to a DC charge current via the circular photogalvanic effect \cite{asnin1979circular,ganichev2001conversion,he2007circular,ogawa2014photocontrol}.  As in the Rashba-Edelstein effect, a nonequilibrium spin accumulation also results from this distribution \cite{edelstein1990spin}, and the nonequilibrium spin exerts a torque on the magnetization.  Our results show that under certain system configurations, Rashba spin-orbit coupling can strongly influence the direction the optical torque.

\section{Model}

\subsection{System description}
Our starting point is an effective model for a Rashba semiconductor with perovskite lattice structure \cite{kim2014switchable}.  This is a convenient model for studying optical transitions with Rashba spin-orbit coupling, and describes recently studied mixed halide perovskite semiconductors which exhibit both exceptional optical absorption and strong spin-orbit coupling \cite{niesner2016giant}.  In these materials the valence band has a predominantly $s$-like orbital character and consists of spin $S=1/2$ states, while the conduction band consists of the spin-orbit splitoff $J=1/2$ states.  This band ordering is opposite to that of commonly studied semiconductors such as GaAs (see Appendix A for more details about the electronic structure).  In terms of real space atomic orbitals $|p_{x,y,z}\rangle$ and spin $|\uparrow,\downarrow\rangle$, the $J_z=+1/2$ state is given by:
\begin{eqnarray}
|J^{+1/2}_{1/2}\rangle = -\frac{1}{\sqrt{3}}\left(|p_x,\downarrow \rangle + i|p_y,\downarrow\rangle + |p_z,\uparrow\rangle\right).
\end{eqnarray}
Note that the expectation value of the spin is anti-parallel to ${\bf J}$ and has a magnitude of 1/6.

\begin{figure}[!h]
  \includegraphics[scale=0.38]{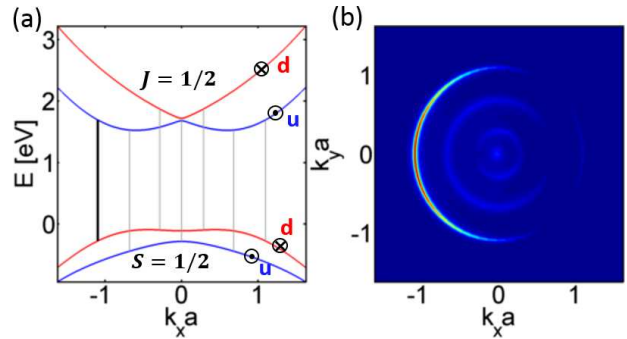}
  \caption{(a) shows the band structure of the perovskite with $\Delta=0.15~{\rm eV}$, $B_R=0.5~{\rm eV}$. Gray vertical lines indicate positions of energetically allowed transitions for $\hbar\omega=1.9~{\rm eV}$, while the thicker black vertical line indicates the dominant transition. $u(d)$ label states aligned (anti-aligned) with the effective magnetic field, and $a$ is the lattice constant.  For $k_x>0$, the $u$ state for conduction (valence) band corresponds to $J_y<0$ ($S_y<0$), denoted with a dot, while the $d$ state is denoted with an x. (b) shows the ${\bf k}$-resolved steady state density upon illumination by light with circular polarization along the $\bf y$-direction.}
    \label{fig:bands}
\end{figure}

Rashba spin-orbit coupling arises from inversion symmetry breaking.  The spin-orbit coupling acts directly on the conduction band $J$ states, and indirectly on the valence band states due to $s$-$p$ hybridization.  To include ferromagnetism we add a spin-dependent exchange field of magnitude $\Delta$.  Due to the spin character of the states described above, the exchange field results in a spin splitting $\Delta^c=\Delta$ of the $S=1/2$ valence band and $\Delta^v=-\Delta/3$ of the $J=1/2$ conduction band.  The Hamiltonian for the system consists of the conduction band $H_c$, the valence band $H_v$, and the conduction-valence band coupling $H_{c-v}$, given by:
\begin{eqnarray}
H_c &=& t_c k^2 + \alpha_c {\bf\sigma} \cdot \left({\bf k}\times{\bf z}\right) - \sigma_z\left( \Delta/2\right)+ \epsilon_0~, \label{eq:Hc}\\
H_v &=&  -t_v k^2 + \alpha_v(k) {\bf \sigma} \cdot \left({\bf k}\times{\bf z}\right) + \sigma_z \left(\Delta/6\right)~, \label{eq:Hv} \\
H_{c-v} &=& i\xi \left({\bf k}\cdot {\bf \sigma}\right) + \gamma_1\left(4-k_+k_-\right)\sigma_z \label{eq:Hvc}.
\end{eqnarray}
Here $\sigma$ is the Pauli spin matrix, $t_{c(v)}$ is the conduction (valence) intraband hopping parameter, $\alpha_{c(v)}$ is the effective Rashba parameter for conduction (valence) band, $\xi$ is the $s$-$p$ interband hopping parameter, $\gamma_1$ is the $s$-$p$ hopping parameter associated with broken inversion symmetry, $\epsilon_0$ is an energy offset for the conduction band, and $k$ is the (dimensionless) Bloch wave vector.  The spin-orbit splitting of the valence band $\alpha_v(k)$ relies on hybridization with the conduction band.  For this reason $\alpha_v(k)$ is generally smaller than the spin-orbit splitting of the conduction band, and varies non-monotonically with $k$ (see Appendix A for the full form of $\alpha_v(k)$).

Figure \ref{fig:bands}(a) shows the model band structure.  We label states parallel (anti-parallel) to the ${\bf k}$-dependent effective magnetic field $u$($d$).  Our default parameters lead to the same sign for the Rashba parameter for conduction and valence bands \cite{footnote2}.  We define $B_R$ as the Rashba-derived splitting at the conduction band minimum.

Appendix A gives the more general form of the $4\times 4$ Hamiltonian in terms of basic tight binding hopping parameters, together with the default values of the parameters used in this work.  Eqs. \ref{eq:Hc}-\ref{eq:Hvc} are a good approximation in the limit where the band gap is greater than other energy scales.  This system represents a minimal model in which to study optical torques, and is amenable to closed form results which elucidate the physics.  We pay a price for this simplicity: some conclusions derived with this model are not directly applicable to materials with more complex electronic structure.  However our analysis provides a framework with which to rationalize the behavior of more realistic systems.  We discuss this more fully in Sec. \ref{sec:discussion}.

We note that spin-orbit coupling enters the model explicitly through $H_{c-v}$, and implicitly via the assumption that the $J=1/2$ band is split-off from the $J=3/2$ bands. The spin-orbit coupling, together with the restriction of ${\bf k}$ to the 2-d plane results in a small magnetic anisotropy in the system, so that an out-of-plane orientation is inequivalent to an in-plane orientation.  Finally, in this study we consider hole-doped system, with Fermi energy $E_F$ sufficiently small so that there are no interband transitions between valence bands.  However we do not self-consistently determine the magnetic exchange splitting $\Delta$ in terms of $E_F$; we take $\Delta$ to be a free parameter.

\subsection{Formalism}

In Appendix B, we derive the formula for the steady state density matrix under monotonic optical excitation ${\bf E}\cos(\omega t)$.  The $j,k$ component of the steady state hole density matrix $\rho^h_{jk}$ is:
\begin{eqnarray}
\rho_{jk}^h({\bf k}) &=& \frac{1}{\epsilon+i\left(E_j^v-E_k^v\right)} \sum_{\ell\in c}\frac{i}{4} \left(\frac{v_{j\ell}v_{k\ell}^*}{\hbar\omega - \left(E_\ell^c-E_j^v\right) - i\epsilon}  \right. \nonumber  \\ && \left.  -\frac{v_{j\ell}^*v_{k\ell}}{\hbar\omega - \left(E_\ell^c-E_k^v\right) + i\epsilon}\right)~, \label{eq:rho}
\end{eqnarray}
where the subscripts $j,k$ refer to the $u,d$ valence bands, the sum $\ell$ is over conduction band states, $\epsilon$ is the $\bf k$-independent broadening associated with the finite lifetime of carriers, and $E_j^{c(v)}$ is the (${\bf k}$ dependent) $j$-th energy eigenvalue of conduction (valence) band.  The dipole transition matrix element $v_{jk}$ is ${v_{jk} = i\langle j | {\bf v} \cdot {\bf E} | k\rangle/\left( E_j^v - E_k^c\right)}$, where ${\bf v} = \frac{\partial H }{ \partial {\bf k}}$.  The dipole transition matrix is determined by the conduction-valence band coupling of Eq. \ref{eq:Hvc}.  In the limit where of small $k$ \footnote{An expansion of interband velocity matrix element in $k$ yields: $v_{jk} = \xi + k^2\xi/E_g^2 \left(2 (t_c+t_v) E_g - 2\xi^2\right)$.  The restriction on $k$ which engures the validity of Eq. \ref{eq:v} is that the second term must be smaller than the first.  For default system parameters, this corresponds to $k<0.5$.}, the velocity operator is given by:
\begin{eqnarray}
v_{jk} \approx -i\xi\frac{\langle j | {\bf \sigma} \cdot {\bf {\bf E}} | k\rangle }{ E_j^v - E_k^c}~ . \label{eq:v}
\end{eqnarray}
We use this approximation in the analytic results of the next section.

Performing the sum over crystal momentum and the trace over the diagonal components of $\rho^h$ gives the steady state, nonthermalized photoexcited hole-density $n$:  \begin{eqnarray}
n = \sum_{\bf k} \left(\rho_{uu}^h({\bf k}) + \rho_{dd}^h({\bf k})\right). \label{eq:phi}
\end{eqnarray}
We find it's useful to present results in terms of the generation rate density $\dot{n}$, which is given by:
\begin{eqnarray}
\dot{n} = \frac{\epsilon }{\hbar}~n.
\end{eqnarray}
The above equation follows from identifying $\hbar/\epsilon$ as the carrier lifetime $\tau$, and noting that in steady state, $n={\dot n} \tau$.

The spin density for holes and electrons is given by:
\begin{eqnarray}
{\bf s}^h &=& +\frac{1}{2}{\rm Tr}\left[\rho^h \sigma\right]~, \\
{\bf s}^e &=& -\frac{1}{6}{\rm Tr}\left[\rho^e \sigma\right]~.
\end{eqnarray}
The relative sign and magnitude of electron and hole spin are derived from the spin of the $J=1/2$ state, as discussed earlier. The torque on the magnetization is determined by the spin component transverse to the magnetization \cite{nunez2006theory,haney2007current}:
\begin{eqnarray}
{\bf \Gamma} = \frac{\Delta}{\hbar}\left({\bf s}^e - {\bf s}^h\right)\times{\bf {\hat M}}~. \label{eq:torque}
\end{eqnarray}
For the numerical results, we consider a 2-dimensional system so that the sum over ${\bf k}$ in Eq. \ref{eq:phi} is restricted to $(k_x,k_y)$.

\section{Results}\label{sec:results}

We present analytic results using the Bloch equations with Eqs. \ref{eq:Hc}-\ref{eq:Hvc}, and \ref{eq:v}, and also present numerical results using the full $4\times 4$ Hamiltonian given in Eq. \ref{eq:4bandH}.

\subsection{$\Delta \gg B_{\rm R}$: Optical spin-transfer torque and inverse Faraday effect}

When Rashba spin-orbit coupling is negligible, we find the semiclassical analysis for the optical spin transfer torque is accurate under certain conditions \cite{fernandez2003optical,nvemec2012experimental}.  In Appendix B, we present the general analytical solution for the optical torque.  In the limit where $\hbar\omega \gg E_g + \Delta$ (so that optical transitions involve states which are far from band edges), the hole spin takes on a particularly simple form in terms of the optical angular momentum ${\bf L}$ and magnetization direction (assumed in the ${\bf z}$-direction):
\begin{eqnarray}
{\bf s}^h = \frac{n}{2}\left(\left({\bf L}\cdot{\bf z}\right){\bf z}+ \frac{\left({\bf z}\times{\bf L}\times{\bf z}\right) + \left(\Delta^v/\epsilon\right) \left({\bf L\times\bf z}\right)}{1+\left(\Delta^v/\epsilon\right)^2} \right). \label{eq:sh}
\end{eqnarray}
The derivation in Appendix B shows that the spin density transverse to the magnetization direction in Eq. \ref{eq:sh} arises from interband coherence.  A similar relation holds for electron spin, with the replacement $\Delta^v \rightarrow \Delta^c$.  Given the electron and hole spin, Eq. \ref{eq:torque} immediately yields the torque per absorption rate $\dot{n}$:
\begin{eqnarray}
\frac{\Gamma_x}{\dot{n}} &=& \frac{\left(\Delta/\epsilon\right)}{2}\left(\frac{1}{1+\left(\Delta/\epsilon\right)^2}-\frac{3}{9+\left(\Delta/\epsilon\right)^2}\right), \label{eq:sttx}\\
\frac{\Gamma_y}{\dot{n}} &=& \frac{\left(\Delta/\epsilon\right)^2}{2}\left(\frac{1}{1+\left(\Delta/\epsilon\right)^2} + \frac{1}{9+\left(\Delta/\epsilon\right)^2}\right) . \label{eq:stt}
\end{eqnarray}

\begin{figure}[!h]
  \includegraphics[scale=0.325]{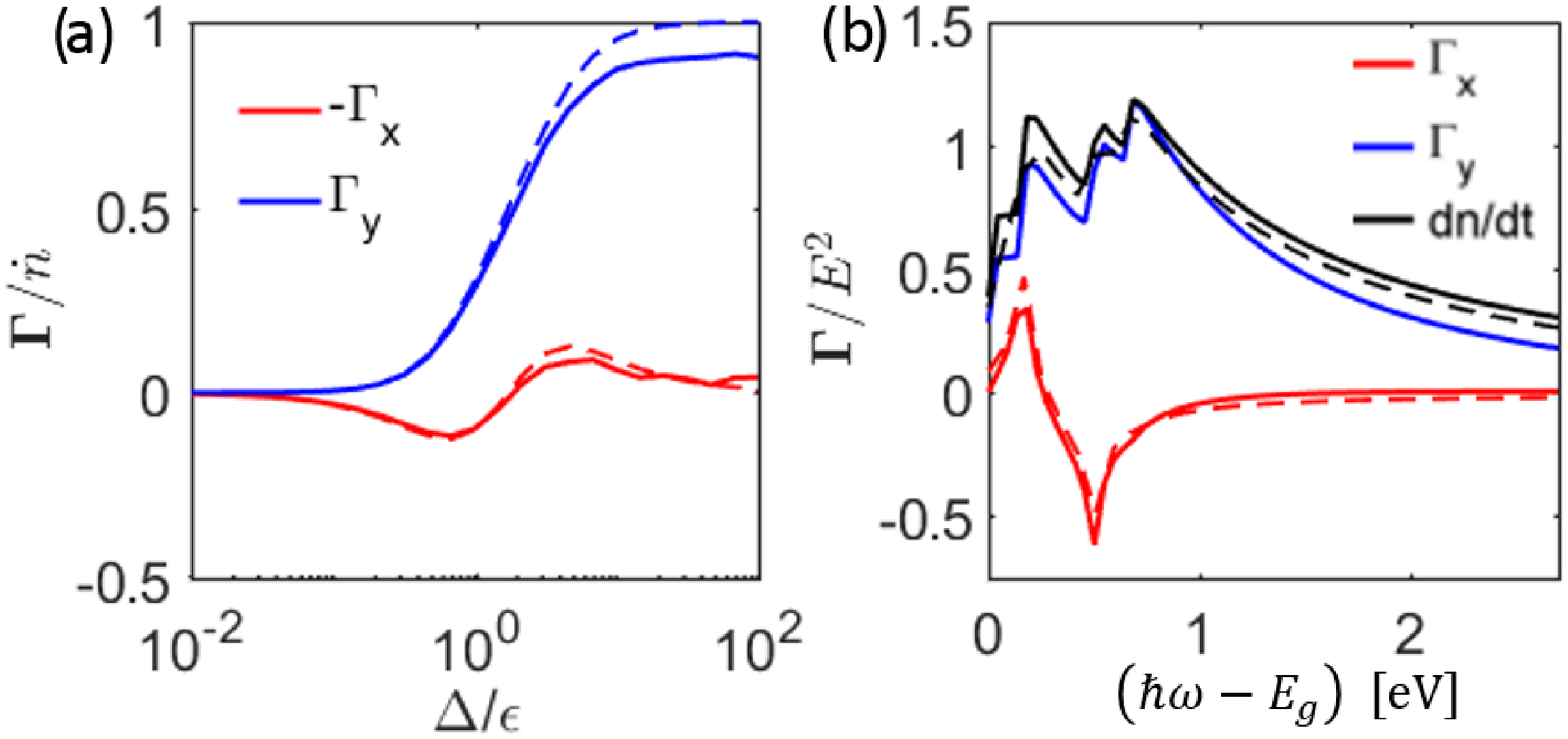}
  \caption{(a) $x$ and $y$ components of the torque per absorption rate as a function of $\Delta/\epsilon$ for $B_R=0$.  System parameters $\epsilon=0.001~{\rm eV}$, the photon energy is $1.9~{\rm eV}$, $\Delta$ is varied between $10^{-5}~{\rm eV}$ to $10^{-1}~{\rm eV}$.  Solid lines are numerically computed values, and dashed lines are Eqs. \ref{eq:sttx}-\ref{eq:stt}. (b) $x,~y$ components of torque, and absorption rate $dn/dt$ versus optical excitation energy.  Solid lines are numerically computed values, and dashed lines are derived from Eqs. \ref{eq:sxfull}-\ref{eq:syfull} for the torque, and Eq. \ref{eq:n} for absorption.}
    \label{fig:stt}
\end{figure}

In the limit $\Delta/\epsilon \gg 1$ we obtain $\Gamma_y=\dot{n}$.  In this case the angular momentum of every absorbed photon is entirely transferred to the magnetization.  In the opposite limit $\Delta/\epsilon \ll 1$, we find that the torque is aligned primarily in the $x$-direction, and is given by $\Gamma_x=\Delta/\left(3\epsilon\right) \dot{n}$.  In this case the absorbed angular momentum is mostly lost to the lattice.  The same result was obtained semi-classically in Ref. \cite{nvemec2012experimental}.  Fig. \ref{fig:stt} shows a comparison of the torques given by Eq. \ref{eq:sttx}-\ref{eq:stt} (given in dashed lines) and the numerical results obtained with the full Hamiltonian (given by solid lines).  We find excellent agreement between the numerical and analytical results.  The discrepancies are due to the approximation of the velocity matrix element.

Fig. \ref{fig:stt}(b) shows the optical torque as a function of excitation energy.  $\Gamma_y$ is approximately equal to $\dot{n}$, as discussed previously, while $\Gamma_x$ exhibits peaks at specific energies.  The full expression for $\Gamma_x$ versus energy is given in Eq. \ref{eq:syfull}, which shows that $\Gamma_x$ is peaked at photon energies corresponding to transitions between near band edge states.  In our model, there are 4 band edge transitions, potentially leading to 4 peaks in $\Gamma_x$.  However two of these peaks in $\Gamma_x$ are suppressed due to cancellation between electron and hole contributions, hence only two peaks are observed.  This torque is derived from the inverse Faraday effect, and we identify its origin as that described in Ref. \cite{qaiumzadeh2013manipulation}, namely a spin-dependent optical stark shift which enables angular momentum to flow between the magnetization and lattice.

\subsection{$\Delta \ll B_{\rm R}$: Optical spin-orbit torque}
When the spin-orbit splitting is greater than the magnetic exchange splitting, the spinors of conduction and valence bands are aligned to the ${\bf k}$-dependent effective magnetic field, which is directed along ${\bf k}\times{\bf z}$.  Letting ${\bf k} =k \left(\cos\left(\theta\right),\sin\left(\theta\right),0\right)$, the spinors take the following form:
\begin{eqnarray}
\psi_u = \frac{1}{\sqrt{2}}\left(
             \begin{array}{c}
               1 \\
               -i e^{i\theta} \\
             \end{array}
           \right) ~~~
\psi_d = \frac{1}{\sqrt{2}}\left(
             \begin{array}{c}
               1 \\
               i e^{i\theta} \\
             \end{array}
           \right) \label{eq:psiR}
\end{eqnarray}
For light with angle of incidence $\beta$ with respect to the surface normal, the dipole matrix elements are:
\begin{eqnarray}
v_{ud,du}&\propto& \pm i \left(\cos\beta \cos\theta - \sin\theta\right) \nonumber \\
v_{uu,dd}&\propto& \mp i \left(e^{-i\theta/2}\sin\left(\frac{\beta}{2}\right) \pm e^{i\theta/2} \cos\left(\frac{\beta}{2}\right) \right)^2 \label{eq:vR}
\end{eqnarray}

In the previous case with negligible Rashba spin-orbit coupling, the transverse spin density results from interband coherence ({\it i.e.} the off-diagonal components of the density matrix).  However in this case the transverse spin density is the result of the misalignment of the eigenstate spin with the magnetization.  For this reason, the net transverse spin density is determined by the diagonal elements of the density matrix.  Using Eqs. \ref{eq:v} and \ref{eq:vR}, we find the diagonal elements of $\rho^h$ in terms of $\theta$ (the direction of ${\bf k}$) and $\beta$ (the direction of ${\bf L}$):
\begin{eqnarray}
\rho_{uu}^h\left(\theta\right) \propto 1+\sin\beta\cos\theta \label{eq:rhouu1} \\
\rho_{dd}^h\left(\theta\right) \propto 1-\sin\beta\cos\theta\label{eq:rhodd1}
\end{eqnarray}
The $\theta$-dependence of $\rho^h$ leads to an asymmetric distribution in $k$-space, as shown in Fig. \ref{fig:bands}(b), which results in a net spin in the ${\bf y}$-direction.

We next use the diagonal components of the density matrix (Eqs. \ref{eq:rhouu1}-\ref{eq:rhodd1}) to estimate the transverse spin density in terms of the nonequilibrium charge density.  To evaluate the spin density, the sum over $\bf k$ for the density matrix is transformed to an integral over $k$ and $\theta$.  Due to the spin texture of the Rashba model, the net spin polarization is determined by the $\theta$ integral.  The $\theta$ integral for the hole density is \cite{footnote3}:
\begin{eqnarray}
n &\propto& 1/\epsilon\int d\theta \left(\rho_{uu}^h\left(\theta\right)+\rho_{dd}^h\left(\theta\right)\right)=8\pi/\epsilon. \label{eq:nR}
\end{eqnarray}
The $\theta$ integral for $S_y^h$ is:
\begin{eqnarray}
S_y^h &\propto& \frac{1}{2\epsilon} \int d\theta \cos\left(\theta\right)\left(\rho_{uu}^h\left(\theta\right)-\rho_{dd}^h\left(\theta\right)\right)\nonumber \\ &&~~~= \left(2\pi/\epsilon\right)\sin\beta . \label{eq:syR}
\end{eqnarray}
We identify the nonzero spin density of Eq. \ref{eq:syR} as an optical Edelstein effect: an asymmetric-in-${\bf k}$ distribution function leads to a nonzero spin density.  Note that this spin density vanishes for normal angle of incidence.  Eqs. \ref{eq:nR}-\ref{eq:syR} yield the spin polarization due to absorption of circularly polarized light is given by:
\begin{eqnarray}
\Rightarrow \frac{S_y^h}{n} &=& \frac{\sin\beta}{4}~.
\end{eqnarray}
A similar analysis for electrons reveals that $S_y^e/n=1/12$.  The resulting torque on the magnetization is along the $x$-direction with magnitude:
\begin{eqnarray}
\Gamma_x &=& n\Delta \left(\frac{1}{4}-\frac{1}{12}\right)\sin\beta  \nonumber \\
\Rightarrow \frac{\Gamma_x}{\dot{n}} &=& \frac{\left(\Delta/\epsilon\right)}{6}\sin\beta ~.\label{eq:ostt}
\end{eqnarray}

There are qualitative differences between the optical torque for the large Rashba case (Eq. \ref{eq:ostt}) and the previous case without Rashba spin-orbit coupling (Eqs. \ref{eq:sttx}-\ref{eq:stt}).  The first difference is the direction of the torque: for large Rashba spin-orbit, the torque is aligned along the $x$-direction for almost all values of system parameters (e.g. $\epsilon$ and $\hbar\omega$), while for the case without Rashba, the direction of the torque varies with system parameters.  The second difference is the scaling of the torque with $\Delta/\epsilon$: for large Rashba spin-orbit coupling, the torque scales as $\Delta/\epsilon$ times the absorption rate, while for the case without Rashba, the torque generally does not exceed the absorption rate \footnote{For vanishing Rashba spin-orbit coupling, the torque derived from the inverse Faraday effect scales as $\ln\left(\Delta/\epsilon\right)$ at photon energies equal to band edge differences.}.  In the presence of strong Rashba spin-orbit coupling, the optical excitation enables angular momentum to flow between the lattice and magnetization, and the magnitude of angular momentum flow exceeds the angular momentum absorption rate if $(\Delta/\epsilon)>6$ for this model.

\begin{figure}[!h]
  \includegraphics[scale=0.29]{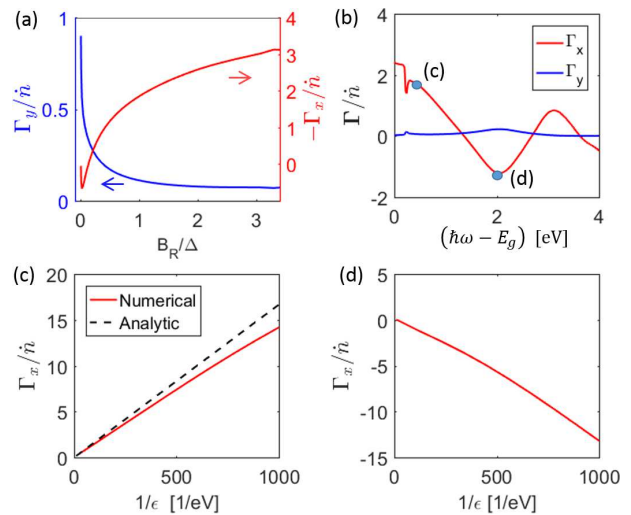}
  \caption{(a) Torque per absorption rate as a function of $B_{\rm R}/\Delta$. For these results, $\epsilon=0.005~{\rm eV}$, $\Delta=0.1~{\rm eV}$, and $\gamma_2$ is varied between 0 eV and -0.3 eV (corresponding to varying $\alpha_c$ between ${(0-0.11)~{\rm eV\cdot nm}}$), and $\gamma_1=1.25\gamma_2$. (b) Torque per absorption rate as a function of excitation energy for default parameters and $\Delta=0.1~{\rm eV}$, $\epsilon=0.008~{\rm eV}$. The (c) and (d) labels on the curve indicate the excitation energies of subplots (c) and (d) of this figure.  (c) $\Gamma_x$ per absorption rate versus $1/\epsilon$, for excitation energy $\hbar\omega-E_g =0.34~{\rm eV}$.  Solid red line is numerical result, dashed black line is Eq. \ref{eq:ostt}.  (d) same as (c), with $\hbar \omega-E_g = 2~{\rm eV}$.}
    \label{fig:results2}
\end{figure}

Figure \ref{fig:results2}(a) shows the crossover between regimes $\Delta \gg B_R$ and $\Delta \ll B_R$ computed numerically, for the case of $\beta=\pi/2$ (corresponding to light polarized along the $y$-direction).  For small $B_R/\Delta$ and large $\Delta/\epsilon$, the torque is along the $y$-direction and equals the absorption rate $\dot{n}$, as discussed in the previous section.  For larger $B_R/\Delta$, the torque is along the $x$-direction and its magnitude exceeds the angular momentum absorption rate.

\begin{figure}[!h]
  \includegraphics[scale=0.35]{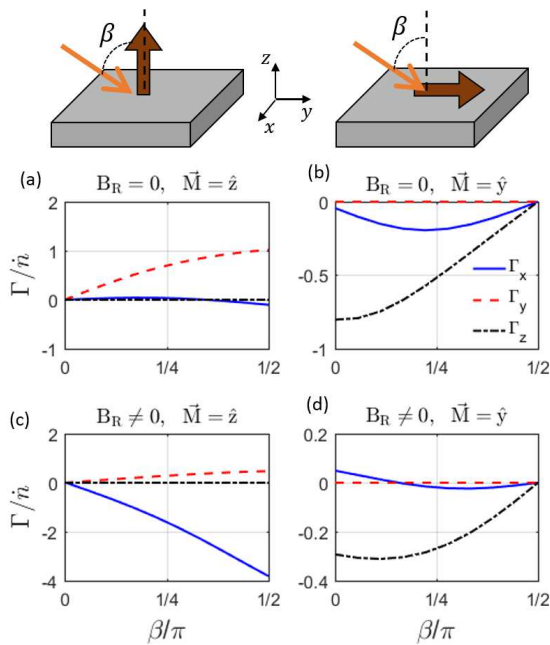}
  \caption{(a) Torque per absorption rate as a function of incoming angle of incidence $\beta$, for no Rashba spin-orbit coupling, and out-of-plane magnetization.  (b) the same data with in-plane magnetization.  (c)  Torque per absorption rate versus $\beta$ with Rashba spin-orbit coupling and out-of-plane magnetization.  (d) the same data with in-plane magnetization.  In all cases $\hbar\omega-E_g=1.4~{\rm eV}$.  In all cases default parameters are used, with $\Delta=1~{\rm eV}$, $\epsilon=0.01~{\rm eV}$.}
    \label{fig:angular}
\end{figure}

Figure \ref{fig:results2}(b) shows $\Gamma_x$ versus excitation energy for large Rashba spin-orbit coupling.  $\Gamma_x$ varies strongly with excitation energy, because the torque is predominantly from high spin holes, and the sign and magnitude of the effective Rashba parameter of the valence bands $\alpha_v(k)$ depends non-monotonically on $k$ (see discussion in Appendix A), and therefore non-monotonically on excitation energy.  Fig. \ref{fig:results2}(c) shows $\Gamma_x$ as a function of inverse broadening $1/\epsilon$ (note $\hbar/\epsilon$ corresponds to the carrier lifetime) for $\hbar\omega-E_g =0.34~{\rm eV}$.  At this excitation energy the assumptions leading to Eq. \ref{eq:ostt} are satisfied (namely, the sign and large magnitude of $B_{\rm R}/\Delta$ for the valence band), and there is good agreement between Eq. \ref{eq:ostt} and numerical results.  Fig. \ref{fig:results2}(d) shows $\Gamma_x$ versus $1/\epsilon$ for $\hbar \omega-E_g = 2~{\rm eV}$, an energy for which the previous analysis doesn't fully apply.  For both values of excitation energy, however, the torque per absorption rate is proportional to $1/\epsilon$.  The torque is not bounded by the absorption rate, but rather by the carrier lifetime.

Finally we present the dependence of the optical torques on the optical and magnetic orientations.  Fig. \ref{fig:angular} shows the numerically computed torque versus angle of incidence $\beta$ for both in-plane and out-of-plane magnetization orientation, with strong and weak Rashba spin-orbit coupling.  For out-of-plane magnetization (panels (a) and (c)), the direction of the optical torque changes with the additional of Rashba spin-orbit, as previously discussed.  For in-plane magnetization (panels (b) and (d)), the torque is aligned along $\Gamma_z$ (the ${\bf M}\times{\bf M}\times{\bf L}$ direction), both in the presence and absence of Rashba spin-orbit coupling.  In this case, the Rashba spin-orbit coupling does not change the direction of the torque because the spin of associated with the optical Edelstein effect is aligned to ${\bf M}$, and therefore doesn't exert a torque.  The torque results instead from the transverse spin density generated from interband coherence.

\section{Discussion}
\label{sec:discussion}

We first comment on the generality of our results, first considering the case of no Rashba spin-orbit coupling.
The generality of our conclusions are limited due to the model system's simple band structure.  With only a single pair of spin-split conduction and valence bands, the magnitude of the band splitting is fixed by the magnetic exchange splitting $\Delta$.  The interband coherence is then determined by a single parameter, $\Delta/\epsilon$.  The transverse spin density and optical torque are in turn expressed with this single parameter.  For realistic band structures, there are multiple spin-opposite pairs of conduction (or valence) bands.  Each pair has its own energy splitting, so the interband coherence and ensuing transverse spin density and torque are not described by a single parameter.  Nevertheless, the size and direction of the transverse spin are set by the spin off-diagonal elements for the density matrix (Eq. (\ref{eq:rho})).  The factors entering this quantity can all be understood in terms of the properties of the band structure and wave functions at a given ${\bf k}$-point.

We next consider the generality of the analysis for the case of strong Rashba spin-orbit coupling and out-of-plane magnetization.  A primary conclusion is that for out-of-plane magnetization and large Rashba splitting (compared to magnetic exchange splitting), the torque is predominantly along the ${\bf M}\times{\bf L}$ direction and is due to an optical Edelstein effect.  This is a more robust conclusion because it does not rely on specifics of our model system.  The torque arises from the misalignment of the eigenstate spin with the magnetization and general optical selection rules.  Unlike the case without Rashba spin-orbit coupling, the torque does not depend on interband coherence and details of the electronic structure.

To provide a feel for the magnitude of the optical torques we compute, we estimate the required photon flux and fluence necessary to induce magnetic switching of a thin film ferromagnet with strong Rashba spin-orbit coupling.  The optically induced spin transfer torque competes with the intrinsic damping torque of the magnetic layer.  For a layer of thickness $t$, out-of-plane anisotropy field $B$, and magnetization $M_s$, the damping rate is $\alpha_d \gamma B t M_s/\mu_B$.  Here $\alpha_d$ is the magnetic damping, $\gamma$ is the gyromagnetic ratio, and $\mu_B$ is is the Bohr magneton.  Setting the optical spin-orbit torque equal to the damping torque and solving for $\Phi$ results in:
\begin{eqnarray}
\Phi = \alpha_d \gamma B t \frac{M_s}{\mu_B} \frac{6\hbar}{\left(\Delta/\epsilon\right)} \frac{1}{W\alpha}~. \label{eq:estimate}
\end{eqnarray}
Here $W$ is the thickness of the absorbing layer, $\alpha$ is the absorption coefficient, and we assume $W\alpha \ll 1$.  For parameter values of $\alpha_d=0.01$, $M_s=10^5~{\rm A/m}$, $B=0.1~{\rm T}$, $\Delta=0.5~{\rm eV}$, $\epsilon=65~{\rm meV}$, $W=5~{\rm nm}$, $t=W$, $\alpha=(100~{\rm nm})^{-1}$, we obtain a value of $\Phi=1.5\times 10^{29}~{\rm m^{-2}\cdot s^{-1}}$.  Choosing an optical pulse length of $1~{\rm ns}$ and a photon energy of $\hbar\omega=1.5~{\rm eV}$, the corresponding fluence is $3.5~{\rm m J/cm^{2}}$.  This can be compared to a maximum fluence of $1~{\rm mJ/cm^2}$ used in Refs. \cite{huisman2016femtosecond,choi2017optical,gorchon2017single}.  This indicates that the influence of Rashba spin-orbit coupling is non-negligible under reasonable assumptions.  A rough estimate for the temperature increase $\Delta T$ from optical absorption is provided via the relation: $\Delta T = \Delta E / C_v$, where $\Delta E$ is the fluence, and $C_v$ is the heat capacity per area of the thin layer (given by the bulk heat capacity multiplied by layer thickness).  For $C_v = 1.9\times 10^{-2}~{\rm J/\left(K\cdot m^2\right)}$, we obtain $\Delta T = 190~{\rm K}$.  This large temperature increase underscores the importance of thermal effects in interpreting experimental results.

There are important extensions of this model which will be considered in future work, such as the inclusion of time-dependence and nonlinear effects \cite{chovan2006ultrafast}.  Experimentally, optical excitation takes the form of the short ($<1~{\rm ps}$), high intensity ($\approx 1~{\rm J/m^2}$ fluence) pulses, so that these effects may be dominant.
Nevertheless the present work provides some basis for intuitively understanding should assist in forming an understanding of more complex conditions.


\acknowledgments{J. L. acknowledges support under the Cooperative Research Agreement between the University of Maryland and the National Institute of Standards and Technology Center for Nanoscale Science and Technology, Award 70NANB10H193, through the University of Maryland.}

\bibliographystyle{apsrev}
\bibliography{ref}

\begin{thebibliography}{44}
\expandafter\ifx\csname natexlab\endcsname\relax\def\natexlab#1{#1}\fi
\expandafter\ifx\csname bibnamefont\endcsname\relax
  \def\bibnamefont#1{#1}\fi
\expandafter\ifx\csname bibfnamefont\endcsname\relax
  \def\bibfnamefont#1{#1}\fi
\expandafter\ifx\csname citenamefont\endcsname\relax
  \def\citenamefont#1{#1}\fi
\expandafter\ifx\csname url\endcsname\relax
  \def\url#1{\texttt{#1}}\fi
\expandafter\ifx\csname urlprefix\endcsname\relax\def\urlprefix{URL }\fi
\providecommand{\bibinfo}[2]{#2}
\providecommand{\eprint}[2][]{\url{#2}}

\bibitem[{\citenamefont{Kirilyuk et~al.}(2010)\citenamefont{Kirilyuk, Kimel,
  and Rasing}}]{kirilyuk2010ultrafast}
\bibinfo{author}{\bibfnamefont{A.}~\bibnamefont{Kirilyuk}},
  \bibinfo{author}{\bibfnamefont{A.~V.} \bibnamefont{Kimel}}, \bibnamefont{and}
  \bibinfo{author}{\bibfnamefont{T.}~\bibnamefont{Rasing}},
  \bibinfo{journal}{Rev. Mod. Phys.} \textbf{\bibinfo{volume}{82}},
  \bibinfo{pages}{2731} (\bibinfo{year}{2010}).

\bibitem[{\citenamefont{Zhang and H{\"u}bner}(2000)}]{zhang2000laser}
\bibinfo{author}{\bibfnamefont{G.~P.} \bibnamefont{Zhang}} \bibnamefont{and}
  \bibinfo{author}{\bibfnamefont{W.}~\bibnamefont{H{\"u}bner}},
  \bibinfo{journal}{Phys. Rev. Lett.} \textbf{\bibinfo{volume}{85}},
  \bibinfo{pages}{3025} (\bibinfo{year}{2000}).

\bibitem[{\citenamefont{Lambert et~al.}(2014)\citenamefont{Lambert, Mangin,
  Varaprasad, Takahashi, Hehn, Cinchetti, Malinowski, Hono, Fainman,
  Aeschlimann et~al.}}]{lambert2014all}
\bibinfo{author}{\bibfnamefont{C.-H.} \bibnamefont{Lambert}},
  \bibinfo{author}{\bibfnamefont{S.}~\bibnamefont{Mangin}},
  \bibinfo{author}{\bibfnamefont{B.~S. D. C.~S.} \bibnamefont{Varaprasad}},
  \bibinfo{author}{\bibfnamefont{Y.}~\bibnamefont{Takahashi}},
  \bibinfo{author}{\bibfnamefont{M.}~\bibnamefont{Hehn}},
  \bibinfo{author}{\bibfnamefont{M.}~\bibnamefont{Cinchetti}},
  \bibinfo{author}{\bibfnamefont{G.}~\bibnamefont{Malinowski}},
  \bibinfo{author}{\bibfnamefont{K.}~\bibnamefont{Hono}},
  \bibinfo{author}{\bibfnamefont{Y.}~\bibnamefont{Fainman}},
  \bibinfo{author}{\bibfnamefont{M.}~\bibnamefont{Aeschlimann}},
  \bibnamefont{et~al.}, \bibinfo{journal}{Science}
  \textbf{\bibinfo{volume}{345}}, \bibinfo{pages}{1337} (\bibinfo{year}{2014}).

\bibitem[{\citenamefont{Guidoni et~al.}(2002)\citenamefont{Guidoni,
  Beaurepaire, and Bigot}}]{guidoni2002magneto}
\bibinfo{author}{\bibfnamefont{L.}~\bibnamefont{Guidoni}},
  \bibinfo{author}{\bibfnamefont{E.}~\bibnamefont{Beaurepaire}},
  \bibnamefont{and} \bibinfo{author}{\bibfnamefont{J.-Y.} \bibnamefont{Bigot}},
  \bibinfo{journal}{Phys. Rev. Lett.} \textbf{\bibinfo{volume}{89}},
  \bibinfo{pages}{017401} (\bibinfo{year}{2002}).

\bibitem[{\citenamefont{Battiato et~al.}(2010)\citenamefont{Battiato, Carva,
  and Oppeneer}}]{battiato2010superdiffusive}
\bibinfo{author}{\bibfnamefont{M.}~\bibnamefont{Battiato}},
  \bibinfo{author}{\bibfnamefont{K.}~\bibnamefont{Carva}}, \bibnamefont{and}
  \bibinfo{author}{\bibfnamefont{P.~M.} \bibnamefont{Oppeneer}},
  \bibinfo{journal}{Phys. Rev. Lett.} \textbf{\bibinfo{volume}{105}},
  \bibinfo{pages}{027203} (\bibinfo{year}{2010}).

\bibitem[{\citenamefont{Duong et~al.}(2004)\citenamefont{Duong, Satoh, and
  Fiebig}}]{duong2004ultrafast}
\bibinfo{author}{\bibfnamefont{N.~P.} \bibnamefont{Duong}},
  \bibinfo{author}{\bibfnamefont{T.}~\bibnamefont{Satoh}}, \bibnamefont{and}
  \bibinfo{author}{\bibfnamefont{M.}~\bibnamefont{Fiebig}},
  \bibinfo{journal}{Phys. Rev. Lett.} \textbf{\bibinfo{volume}{93}},
  \bibinfo{pages}{117402} (\bibinfo{year}{2004}).

\bibitem[{\citenamefont{Hashimoto et~al.}(2008)\citenamefont{Hashimoto,
  Kobayashi, and Munekata}}]{hashimoto2008photoinduced}
\bibinfo{author}{\bibfnamefont{Y.}~\bibnamefont{Hashimoto}},
  \bibinfo{author}{\bibfnamefont{S.}~\bibnamefont{Kobayashi}},
  \bibnamefont{and} \bibinfo{author}{\bibfnamefont{H.}~\bibnamefont{Munekata}},
  \bibinfo{journal}{Phys. Rev. Lett.} \textbf{\bibinfo{volume}{100}},
  \bibinfo{pages}{067202} (\bibinfo{year}{2008}).

\bibitem[{\citenamefont{Tesa{\v{r}}ov{\'a}
  et~al.}(2013)\citenamefont{Tesa{\v{r}}ov{\'a}, N{\v{e}}mec, Rozkotov{\'a},
  Zemen, Janda, Butkovi{\v{c}}ov{\'a}, Troj{\'a}nek, Olejn{\'\i}k, Nov{\'a}k,
  Mal{\`y} et~al.}}]{tesavrova2013experimental}
\bibinfo{author}{\bibfnamefont{N.}~\bibnamefont{Tesa{\v{r}}ov{\'a}}},
  \bibinfo{author}{\bibfnamefont{P.}~\bibnamefont{N{\v{e}}mec}},
  \bibinfo{author}{\bibfnamefont{E.}~\bibnamefont{Rozkotov{\'a}}},
  \bibinfo{author}{\bibfnamefont{J.}~\bibnamefont{Zemen}},
  \bibinfo{author}{\bibfnamefont{T.}~\bibnamefont{Janda}},
  \bibinfo{author}{\bibfnamefont{D.}~\bibnamefont{Butkovi{\v{c}}ov{\'a}}},
  \bibinfo{author}{\bibfnamefont{F.}~\bibnamefont{Troj{\'a}nek}},
  \bibinfo{author}{\bibfnamefont{K.}~\bibnamefont{Olejn{\'\i}k}},
  \bibinfo{author}{\bibfnamefont{V.}~\bibnamefont{Nov{\'a}k}},
  \bibinfo{author}{\bibfnamefont{P.}~\bibnamefont{Mal{\`y}}},
  \bibnamefont{et~al.}, \bibinfo{journal}{Nat. Phot.}
  \textbf{\bibinfo{volume}{7}}, \bibinfo{pages}{492} (\bibinfo{year}{2013}).

\bibitem[{\citenamefont{Fern{\'a}ndez-Rossier
  et~al.}(2003)\citenamefont{Fern{\'a}ndez-Rossier, N{\'u}{\~n}ez, Abolfath,
  and MacDonald}}]{fernandez2003optical}
\bibinfo{author}{\bibfnamefont{J.}~\bibnamefont{Fern{\'a}ndez-Rossier}},
  \bibinfo{author}{\bibfnamefont{A.~S.} \bibnamefont{N{\'u}{\~n}ez}},
  \bibinfo{author}{\bibfnamefont{M.}~\bibnamefont{Abolfath}}, \bibnamefont{and}
  \bibinfo{author}{\bibfnamefont{A.~H.} \bibnamefont{MacDonald}},
  \bibinfo{journal}{arXiv preprint cond-mat/0304492}  (\bibinfo{year}{2003}).

\bibitem[{\citenamefont{Chovan et~al.}(2006)\citenamefont{Chovan, Kavousanaki,
  and Perakis}}]{chovan2006ultrafast}
\bibinfo{author}{\bibfnamefont{J.}~\bibnamefont{Chovan}},
  \bibinfo{author}{\bibfnamefont{E.~G.} \bibnamefont{Kavousanaki}},
  \bibnamefont{and} \bibinfo{author}{\bibfnamefont{I.~E.}
  \bibnamefont{Perakis}}, \bibinfo{journal}{Phys. Rev. Lett.}
  \textbf{\bibinfo{volume}{96}}, \bibinfo{pages}{057402}
  (\bibinfo{year}{2006}).

\bibitem[{\citenamefont{N{\v{e}}mec et~al.}(2012)\citenamefont{N{\v{e}}mec,
  Rozkotov{\'a}, Tesa{\v{r}}ov{\'a}, Troj{\'a}nek, De~Ranieri, Olejn{\'\i}k,
  Zemen, Nov{\'a}k, Cukr, Mal{\`y} et~al.}}]{nvemec2012experimental}
\bibinfo{author}{\bibfnamefont{P.}~\bibnamefont{N{\v{e}}mec}},
  \bibinfo{author}{\bibfnamefont{E.}~\bibnamefont{Rozkotov{\'a}}},
  \bibinfo{author}{\bibfnamefont{N.}~\bibnamefont{Tesa{\v{r}}ov{\'a}}},
  \bibinfo{author}{\bibfnamefont{F.}~\bibnamefont{Troj{\'a}nek}},
  \bibinfo{author}{\bibfnamefont{E.}~\bibnamefont{De~Ranieri}},
  \bibinfo{author}{\bibfnamefont{K.}~\bibnamefont{Olejn{\'\i}k}},
  \bibinfo{author}{\bibfnamefont{J.}~\bibnamefont{Zemen}},
  \bibinfo{author}{\bibfnamefont{V.}~\bibnamefont{Nov{\'a}k}},
  \bibinfo{author}{\bibfnamefont{M.}~\bibnamefont{Cukr}},
  \bibinfo{author}{\bibfnamefont{P.}~\bibnamefont{Mal{\`y}}},
  \bibnamefont{et~al.}, \bibinfo{journal}{Nat. Phys.}
  \textbf{\bibinfo{volume}{8}}, \bibinfo{pages}{411} (\bibinfo{year}{2012}).

\bibitem[{\citenamefont{Slonczewski}(1996)}]{slonczewski1996current}
\bibinfo{author}{\bibfnamefont{J.~C.} \bibnamefont{Slonczewski}},
  \bibinfo{journal}{J. Magn. Magn. Mat.} \textbf{\bibinfo{volume}{159}},
  \bibinfo{pages}{L1} (\bibinfo{year}{1996}).

\bibitem[{\citenamefont{Berger}(1996)}]{berger1996emission}
\bibinfo{author}{\bibfnamefont{L.}~\bibnamefont{Berger}},
  \bibinfo{journal}{Phys. Rev. B} \textbf{\bibinfo{volume}{54}},
  \bibinfo{pages}{9353} (\bibinfo{year}{1996}).

\bibitem[{\citenamefont{Ralph and Stiles}(2008)}]{ralph2008spin}
\bibinfo{author}{\bibfnamefont{D.~C.} \bibnamefont{Ralph}} \bibnamefont{and}
  \bibinfo{author}{\bibfnamefont{M.~D.} \bibnamefont{Stiles}},
  \bibinfo{journal}{J. Magn. Magn. Mat.} \textbf{\bibinfo{volume}{320}},
  \bibinfo{pages}{1190} (\bibinfo{year}{2008}).

\bibitem[{\citenamefont{Haney and Stiles}(2010)}]{haney2010current}
\bibinfo{author}{\bibfnamefont{P.~M.} \bibnamefont{Haney}} \bibnamefont{and}
  \bibinfo{author}{\bibfnamefont{M.~D.} \bibnamefont{Stiles}},
  \bibinfo{journal}{Phys. Rev. Lett.} \textbf{\bibinfo{volume}{105}},
  \bibinfo{pages}{126602} (\bibinfo{year}{2010}).

\bibitem[{\citenamefont{Qaiumzadeh et~al.}(2013)\citenamefont{Qaiumzadeh,
  Bauer, and Brataas}}]{qaiumzadeh2013manipulation}
\bibinfo{author}{\bibfnamefont{A.}~\bibnamefont{Qaiumzadeh}},
  \bibinfo{author}{\bibfnamefont{G.~E.~W.} \bibnamefont{Bauer}},
  \bibnamefont{and} \bibinfo{author}{\bibfnamefont{A.}~\bibnamefont{Brataas}},
  \bibinfo{journal}{Physical Review B} \textbf{\bibinfo{volume}{88}},
  \bibinfo{pages}{064416} (\bibinfo{year}{2013}).

\bibitem[{\citenamefont{Miron et~al.}(2011)\citenamefont{Miron, Garello,
  Gaudin, Zermatten, Costache, Auffret, Bandiera, Rodmacq, Schuhl, and
  Gambardella}}]{miron2011perpendicular}
\bibinfo{author}{\bibfnamefont{I.~M.} \bibnamefont{Miron}},
  \bibinfo{author}{\bibfnamefont{K.}~\bibnamefont{Garello}},
  \bibinfo{author}{\bibfnamefont{G.}~\bibnamefont{Gaudin}},
  \bibinfo{author}{\bibfnamefont{P.-J.} \bibnamefont{Zermatten}},
  \bibinfo{author}{\bibfnamefont{M.~V.} \bibnamefont{Costache}},
  \bibinfo{author}{\bibfnamefont{S.}~\bibnamefont{Auffret}},
  \bibinfo{author}{\bibfnamefont{S.}~\bibnamefont{Bandiera}},
  \bibinfo{author}{\bibfnamefont{B.}~\bibnamefont{Rodmacq}},
  \bibinfo{author}{\bibfnamefont{A.}~\bibnamefont{Schuhl}}, \bibnamefont{and}
  \bibinfo{author}{\bibfnamefont{P.}~\bibnamefont{Gambardella}},
  \bibinfo{journal}{Nature} \textbf{\bibinfo{volume}{476}},
  \bibinfo{pages}{189} (\bibinfo{year}{2011}).

\bibitem[{\citenamefont{Garello et~al.}(2013)\citenamefont{Garello, Miron,
  Avci, Freimuth, Mokrousov, Bl{\"u}gel, Auffret, Boulle, Gaudin, and
  Gambardella}}]{garello2013symmetry}
\bibinfo{author}{\bibfnamefont{K.}~\bibnamefont{Garello}},
  \bibinfo{author}{\bibfnamefont{I.~M.} \bibnamefont{Miron}},
  \bibinfo{author}{\bibfnamefont{C.~O.} \bibnamefont{Avci}},
  \bibinfo{author}{\bibfnamefont{F.}~\bibnamefont{Freimuth}},
  \bibinfo{author}{\bibfnamefont{Y.}~\bibnamefont{Mokrousov}},
  \bibinfo{author}{\bibfnamefont{S.}~\bibnamefont{Bl{\"u}gel}},
  \bibinfo{author}{\bibfnamefont{S.}~\bibnamefont{Auffret}},
  \bibinfo{author}{\bibfnamefont{O.}~\bibnamefont{Boulle}},
  \bibinfo{author}{\bibfnamefont{G.}~\bibnamefont{Gaudin}}, \bibnamefont{and}
  \bibinfo{author}{\bibfnamefont{P.}~\bibnamefont{Gambardella}},
  \bibinfo{journal}{Nat. Nanotech.} \textbf{\bibinfo{volume}{8}},
  \bibinfo{pages}{587} (\bibinfo{year}{2013}).

\bibitem[{\citenamefont{Kim et~al.}(2013)\citenamefont{Kim, Sinha, Hayashi,
  Yamanouchi, Fukami, Suzuki, Mitani, and Ohno}}]{kim2013layer}
\bibinfo{author}{\bibfnamefont{J.}~\bibnamefont{Kim}},
  \bibinfo{author}{\bibfnamefont{J.}~\bibnamefont{Sinha}},
  \bibinfo{author}{\bibfnamefont{M.}~\bibnamefont{Hayashi}},
  \bibinfo{author}{\bibfnamefont{M.}~\bibnamefont{Yamanouchi}},
  \bibinfo{author}{\bibfnamefont{S.}~\bibnamefont{Fukami}},
  \bibinfo{author}{\bibfnamefont{T.}~\bibnamefont{Suzuki}},
  \bibinfo{author}{\bibfnamefont{S.}~\bibnamefont{Mitani}}, \bibnamefont{and}
  \bibinfo{author}{\bibfnamefont{H.}~\bibnamefont{Ohno}},
  \bibinfo{journal}{Nat. Mat.} \textbf{\bibinfo{volume}{12}},
  \bibinfo{pages}{240} (\bibinfo{year}{2013}).

\bibitem[{\citenamefont{Qaiumzadeh et~al.}(2015)\citenamefont{Qaiumzadeh,
  Duine, and Titov}}]{qaiumzadeh2015spin}
\bibinfo{author}{\bibfnamefont{A.}~\bibnamefont{Qaiumzadeh}},
  \bibinfo{author}{\bibfnamefont{R.~A.} \bibnamefont{Duine}}, \bibnamefont{and}
  \bibinfo{author}{\bibfnamefont{M.}~\bibnamefont{Titov}},
  \bibinfo{journal}{Physical Review B} \textbf{\bibinfo{volume}{92}},
  \bibinfo{pages}{014402} (\bibinfo{year}{2015}).

\bibitem[{\citenamefont{Liu et~al.}(2012)\citenamefont{Liu, Pai, Li, Tseng,
  Ralph, and Buhrman}}]{liu2012spin}
\bibinfo{author}{\bibfnamefont{L.}~\bibnamefont{Liu}},
  \bibinfo{author}{\bibfnamefont{C.-F.} \bibnamefont{Pai}},
  \bibinfo{author}{\bibfnamefont{Y.}~\bibnamefont{Li}},
  \bibinfo{author}{\bibfnamefont{H.~W.} \bibnamefont{Tseng}},
  \bibinfo{author}{\bibfnamefont{D.~C.} \bibnamefont{Ralph}}, \bibnamefont{and}
  \bibinfo{author}{\bibfnamefont{R.~A.} \bibnamefont{Buhrman}},
  \bibinfo{journal}{Science} \textbf{\bibinfo{volume}{336}},
  \bibinfo{pages}{555} (\bibinfo{year}{2012}).

\bibitem[{\citenamefont{Kapetanakis et~al.}(2009)\citenamefont{Kapetanakis,
  Perakis, Wickey, Piermarocchi, and Wang}}]{kapetanakis2009femtosecond}
\bibinfo{author}{\bibfnamefont{M.~D.} \bibnamefont{Kapetanakis}},
  \bibinfo{author}{\bibfnamefont{I.~E.} \bibnamefont{Perakis}},
  \bibinfo{author}{\bibfnamefont{K.~J.} \bibnamefont{Wickey}},
  \bibinfo{author}{\bibfnamefont{C.}~\bibnamefont{Piermarocchi}},
  \bibnamefont{and} \bibinfo{author}{\bibfnamefont{J.}~\bibnamefont{Wang}},
  \bibinfo{journal}{Physical review letters} \textbf{\bibinfo{volume}{103}},
  \bibinfo{pages}{047404} (\bibinfo{year}{2009}).

\bibitem[{\citenamefont{Berritta et~al.}(2016)\citenamefont{Berritta, Mondal,
  Carva, and Oppeneer}}]{berritta2016ab}
\bibinfo{author}{\bibfnamefont{M.}~\bibnamefont{Berritta}},
  \bibinfo{author}{\bibfnamefont{R.}~\bibnamefont{Mondal}},
  \bibinfo{author}{\bibfnamefont{K.}~\bibnamefont{Carva}}, \bibnamefont{and}
  \bibinfo{author}{\bibfnamefont{P.~M.} \bibnamefont{Oppeneer}},
  \bibinfo{journal}{Physical Review Letters} \textbf{\bibinfo{volume}{117}},
  \bibinfo{pages}{137203} (\bibinfo{year}{2016}).

\bibitem[{\citenamefont{Freimuth et~al.}(2016)\citenamefont{Freimuth,
  Bl{\"u}gel, and Mokrousov}}]{freimuth2016laser}
\bibinfo{author}{\bibfnamefont{F.}~\bibnamefont{Freimuth}},
  \bibinfo{author}{\bibfnamefont{S.}~\bibnamefont{Bl{\"u}gel}},
  \bibnamefont{and}
  \bibinfo{author}{\bibfnamefont{Y.}~\bibnamefont{Mokrousov}},
  \bibinfo{journal}{Phys. Rev. B} \textbf{\bibinfo{volume}{94}},
  \bibinfo{pages}{144432} (\bibinfo{year}{2016}).

\bibitem[{\citenamefont{Qaiumzadeh and Titov}(2016)}]{qaiumzadeh2016theory}
\bibinfo{author}{\bibfnamefont{A.}~\bibnamefont{Qaiumzadeh}} \bibnamefont{and}
  \bibinfo{author}{\bibfnamefont{M.}~\bibnamefont{Titov}},
  \bibinfo{journal}{Phys. Rev. B} \textbf{\bibinfo{volume}{94}},
  \bibinfo{pages}{014425} (\bibinfo{year}{2016}).

\bibitem[{\citenamefont{Gorchon et~al.}(2017)\citenamefont{Gorchon, Lambert,
  Yang, Pattabi, Wilson, Salahuddin, and Bokor}}]{gorchon2017single}
\bibinfo{author}{\bibfnamefont{J.}~\bibnamefont{Gorchon}},
  \bibinfo{author}{\bibfnamefont{C.-H.} \bibnamefont{Lambert}},
  \bibinfo{author}{\bibfnamefont{Y.}~\bibnamefont{Yang}},
  \bibinfo{author}{\bibfnamefont{A.}~\bibnamefont{Pattabi}},
  \bibinfo{author}{\bibfnamefont{R.~B.} \bibnamefont{Wilson}},
  \bibinfo{author}{\bibfnamefont{S.}~\bibnamefont{Salahuddin}},
  \bibnamefont{and} \bibinfo{author}{\bibfnamefont{J.}~\bibnamefont{Bokor}},
  \bibinfo{journal}{arXiv preprint arXiv:1702.08491}  (\bibinfo{year}{2017}).

\bibitem[{\citenamefont{Choi et~al.}(2017)\citenamefont{Choi, Schleife, and
  Cahill}}]{choi2017optical}
\bibinfo{author}{\bibfnamefont{G.-M.} \bibnamefont{Choi}},
  \bibinfo{author}{\bibfnamefont{A.}~\bibnamefont{Schleife}}, \bibnamefont{and}
  \bibinfo{author}{\bibfnamefont{D.~G.} \bibnamefont{Cahill}},
  \bibinfo{journal}{Nature communications} \textbf{\bibinfo{volume}{8}},
  \bibinfo{pages}{15085} (\bibinfo{year}{2017}).

\bibitem[{\citenamefont{Pershan et~al.}(1966)\citenamefont{Pershan, Van~der
  Ziel, and Malmstrom}}]{pershan1966theoretical}
\bibinfo{author}{\bibfnamefont{P.~S.} \bibnamefont{Pershan}},
  \bibinfo{author}{\bibfnamefont{J.~P.} \bibnamefont{Van~der Ziel}},
  \bibnamefont{and} \bibinfo{author}{\bibfnamefont{L.~D.}
  \bibnamefont{Malmstrom}}, \bibinfo{journal}{Physical Review}
  \textbf{\bibinfo{volume}{143}}, \bibinfo{pages}{574} (\bibinfo{year}{1966}).

\bibitem[{\citenamefont{Kimel et~al.}(2005)\citenamefont{Kimel, Kirilyuk,
  Usachev, Pisarev, Balbashov, and Rasing}}]{kimel2005ultrafast}
\bibinfo{author}{\bibfnamefont{A.~V.} \bibnamefont{Kimel}},
  \bibinfo{author}{\bibfnamefont{A.}~\bibnamefont{Kirilyuk}},
  \bibinfo{author}{\bibfnamefont{P.~A.} \bibnamefont{Usachev}},
  \bibinfo{author}{\bibfnamefont{R.~V.} \bibnamefont{Pisarev}},
  \bibinfo{author}{\bibfnamefont{A.~M.} \bibnamefont{Balbashov}},
  \bibnamefont{and} \bibinfo{author}{\bibfnamefont{T.}~\bibnamefont{Rasing}},
  \bibinfo{journal}{Nature} \textbf{\bibinfo{volume}{435}},
  \bibinfo{pages}{655} (\bibinfo{year}{2005}).

\bibitem[{\citenamefont{Manchon and Zhang}(2008)}]{manchon2008theory}
\bibinfo{author}{\bibfnamefont{A.}~\bibnamefont{Manchon}} \bibnamefont{and}
  \bibinfo{author}{\bibfnamefont{S.}~\bibnamefont{Zhang}},
  \bibinfo{journal}{Phys. Rev. B} \textbf{\bibinfo{volume}{78}},
  \bibinfo{pages}{212405} (\bibinfo{year}{2008}).

\bibitem[{\citenamefont{Garate and MacDonald}(2009)}]{garate2009influence}
\bibinfo{author}{\bibfnamefont{I.}~\bibnamefont{Garate}} \bibnamefont{and}
  \bibinfo{author}{\bibfnamefont{A.~H.} \bibnamefont{MacDonald}},
  \bibinfo{journal}{Phys. Rev. B} \textbf{\bibinfo{volume}{80}},
  \bibinfo{pages}{134403} (\bibinfo{year}{2009}).

\bibitem[{\citenamefont{Asnin et~al.}(1979)\citenamefont{Asnin, Bakun,
  Danishevskii, Ivchenko, Pikus, and Rogachev}}]{asnin1979circular}
\bibinfo{author}{\bibfnamefont{V.~M.} \bibnamefont{Asnin}},
  \bibinfo{author}{\bibfnamefont{A.~A.} \bibnamefont{Bakun}},
  \bibinfo{author}{\bibfnamefont{A.~M.} \bibnamefont{Danishevskii}},
  \bibinfo{author}{\bibfnamefont{E.~L.} \bibnamefont{Ivchenko}},
  \bibinfo{author}{\bibfnamefont{G.~E.} \bibnamefont{Pikus}}, \bibnamefont{and}
  \bibinfo{author}{\bibfnamefont{A.~A.} \bibnamefont{Rogachev}},
  \bibinfo{journal}{Sol. St. Comm.} \textbf{\bibinfo{volume}{30}},
  \bibinfo{pages}{565} (\bibinfo{year}{1979}).

\bibitem[{\citenamefont{Ganichev et~al.}(2001)\citenamefont{Ganichev, Ivchenko,
  Danilov, Eroms, Wegscheider, Weiss, and Prettl}}]{ganichev2001conversion}
\bibinfo{author}{\bibfnamefont{S.~D.} \bibnamefont{Ganichev}},
  \bibinfo{author}{\bibfnamefont{E.~L.} \bibnamefont{Ivchenko}},
  \bibinfo{author}{\bibfnamefont{S.~N.} \bibnamefont{Danilov}},
  \bibinfo{author}{\bibfnamefont{J.}~\bibnamefont{Eroms}},
  \bibinfo{author}{\bibfnamefont{W.}~\bibnamefont{Wegscheider}},
  \bibinfo{author}{\bibfnamefont{D.}~\bibnamefont{Weiss}}, \bibnamefont{and}
  \bibinfo{author}{\bibfnamefont{W.}~\bibnamefont{Prettl}},
  \bibinfo{journal}{Phys. Rev. Lett.} \textbf{\bibinfo{volume}{86}},
  \bibinfo{pages}{4358} (\bibinfo{year}{2001}).

\bibitem[{\citenamefont{He et~al.}(2007)\citenamefont{He, Shen, Tang, Tang,
  Yin, Xu, Yang, Zhang, Chen, Tang et~al.}}]{he2007circular}
\bibinfo{author}{\bibfnamefont{X.~W.} \bibnamefont{He}},
  \bibinfo{author}{\bibfnamefont{B.}~\bibnamefont{Shen}},
  \bibinfo{author}{\bibfnamefont{Y.~Q.} \bibnamefont{Tang}},
  \bibinfo{author}{\bibfnamefont{N.}~\bibnamefont{Tang}},
  \bibinfo{author}{\bibfnamefont{C.~M.} \bibnamefont{Yin}},
  \bibinfo{author}{\bibfnamefont{F.~J.} \bibnamefont{Xu}},
  \bibinfo{author}{\bibfnamefont{Z.~J.} \bibnamefont{Yang}},
  \bibinfo{author}{\bibfnamefont{G.~Y.} \bibnamefont{Zhang}},
  \bibinfo{author}{\bibfnamefont{Y.~H.} \bibnamefont{Chen}},
  \bibinfo{author}{\bibfnamefont{C.~G.} \bibnamefont{Tang}},
  \bibnamefont{et~al.}, \bibinfo{journal}{App. Phys. Lett.}
  \textbf{\bibinfo{volume}{91}}, \bibinfo{pages}{071912}
  (\bibinfo{year}{2007}).

\bibitem[{\citenamefont{Ogawa et~al.}(2014)\citenamefont{Ogawa, Bahramy,
  Kaneko, and Tokura}}]{ogawa2014photocontrol}
\bibinfo{author}{\bibfnamefont{N.}~\bibnamefont{Ogawa}},
  \bibinfo{author}{\bibfnamefont{M.~S.} \bibnamefont{Bahramy}},
  \bibinfo{author}{\bibfnamefont{Y.}~\bibnamefont{Kaneko}}, \bibnamefont{and}
  \bibinfo{author}{\bibfnamefont{Y.}~\bibnamefont{Tokura}},
  \bibinfo{journal}{Phys. Rev. B} \textbf{\bibinfo{volume}{90}},
  \bibinfo{pages}{125122} (\bibinfo{year}{2014}).

\bibitem[{\citenamefont{Edelstein}(1990)}]{edelstein1990spin}
\bibinfo{author}{\bibfnamefont{V.~M.} \bibnamefont{Edelstein}},
  \bibinfo{journal}{Sol. St. Comm.} \textbf{\bibinfo{volume}{73}},
  \bibinfo{pages}{233} (\bibinfo{year}{1990}).

\bibitem[{\citenamefont{Kim et~al.}(2014)\citenamefont{Kim, Im, Freeman, Ihm,
  and Jin}}]{kim2014switchable}
\bibinfo{author}{\bibfnamefont{M.}~\bibnamefont{Kim}},
  \bibinfo{author}{\bibfnamefont{J.}~\bibnamefont{Im}},
  \bibinfo{author}{\bibfnamefont{A.~J.} \bibnamefont{Freeman}},
  \bibinfo{author}{\bibfnamefont{J.}~\bibnamefont{Ihm}}, \bibnamefont{and}
  \bibinfo{author}{\bibfnamefont{H.}~\bibnamefont{Jin}},
  \bibinfo{journal}{Proc. Nat. Ac. Sc.} \textbf{\bibinfo{volume}{111}},
  \bibinfo{pages}{6900} (\bibinfo{year}{2014}).

\bibitem[{\citenamefont{Niesner et~al.}(2016)\citenamefont{Niesner, Wilhelm,
  Levchuk, Osvet, Shrestha, Batentschuk, Brabec, and
  Fauster}}]{niesner2016giant}
\bibinfo{author}{\bibfnamefont{D.}~\bibnamefont{Niesner}},
  \bibinfo{author}{\bibfnamefont{M.}~\bibnamefont{Wilhelm}},
  \bibinfo{author}{\bibfnamefont{I.}~\bibnamefont{Levchuk}},
  \bibinfo{author}{\bibfnamefont{A.}~\bibnamefont{Osvet}},
  \bibinfo{author}{\bibfnamefont{S.}~\bibnamefont{Shrestha}},
  \bibinfo{author}{\bibfnamefont{M.}~\bibnamefont{Batentschuk}},
  \bibinfo{author}{\bibfnamefont{C.}~\bibnamefont{Brabec}}, \bibnamefont{and}
  \bibinfo{author}{\bibfnamefont{T.}~\bibnamefont{Fauster}},
  \bibinfo{journal}{Phys. Rev. Lett.} \textbf{\bibinfo{volume}{117}},
  \bibinfo{pages}{126401} (\bibinfo{year}{2016}).

\bibitem[{foo({\natexlab{a}})}]{footnote2}
\bibinfo{note}{We find that the results are qualitatively similar (differ by
  less than a factor of 2) if $\alpha_c$ and $\alpha_v$ have different signs.}

\bibitem[{\citenamefont{N{\'u}{\~n}ez and MacDonald}(2006)}]{nunez2006theory}
\bibinfo{author}{\bibfnamefont{A.~S.} \bibnamefont{N{\'u}{\~n}ez}}
  \bibnamefont{and} \bibinfo{author}{\bibfnamefont{A.~H.}
  \bibnamefont{MacDonald}}, \bibinfo{journal}{Sol. St. Comm.}
  \textbf{\bibinfo{volume}{139}}, \bibinfo{pages}{31} (\bibinfo{year}{2006}).

\bibitem[{\citenamefont{Haney et~al.}(2007)\citenamefont{Haney, Waldron, Duine,
  Nunez, Guo, and MacDonald}}]{haney2007current}
\bibinfo{author}{\bibfnamefont{P.~M.} \bibnamefont{Haney}},
  \bibinfo{author}{\bibfnamefont{D.}~\bibnamefont{Waldron}},
  \bibinfo{author}{\bibfnamefont{R.~A.} \bibnamefont{Duine}},
  \bibinfo{author}{\bibfnamefont{A.~S.} \bibnamefont{Nunez}},
  \bibinfo{author}{\bibfnamefont{H.}~\bibnamefont{Guo}}, \bibnamefont{and}
  \bibinfo{author}{\bibfnamefont{A.~H.} \bibnamefont{MacDonald}},
  \bibinfo{journal}{Phys. Rev. B} \textbf{\bibinfo{volume}{76}},
  \bibinfo{pages}{024404} (\bibinfo{year}{2007}).

\bibitem[{foo({\natexlab{b}})}]{footnote3}
\bibinfo{note}{In writing this relation, we assume that the band gap is much
  greater than other band splittings, so that $E_{jk}^c-E_{j'k'}^v$ is constant
  for all $j,k,j',k'$.}

\bibitem[{\citenamefont{Huisman et~al.}(2016)\citenamefont{Huisman,
  Mikhaylovskiy, Costa, Freimuth, Paz, Ventura, Freitas, Bl{\"u}gel, Mokrousov,
  Rasing et~al.}}]{huisman2016femtosecond}
\bibinfo{author}{\bibfnamefont{T.~J.} \bibnamefont{Huisman}},
  \bibinfo{author}{\bibfnamefont{R.~V.} \bibnamefont{Mikhaylovskiy}},
  \bibinfo{author}{\bibfnamefont{J.~D.} \bibnamefont{Costa}},
  \bibinfo{author}{\bibfnamefont{F.}~\bibnamefont{Freimuth}},
  \bibinfo{author}{\bibfnamefont{E.}~\bibnamefont{Paz}},
  \bibinfo{author}{\bibfnamefont{J.}~\bibnamefont{Ventura}},
  \bibinfo{author}{\bibfnamefont{P.~P.} \bibnamefont{Freitas}},
  \bibinfo{author}{\bibfnamefont{S.}~\bibnamefont{Bl{\"u}gel}},
  \bibinfo{author}{\bibfnamefont{Y.}~\bibnamefont{Mokrousov}},
  \bibinfo{author}{\bibfnamefont{T.}~\bibnamefont{Rasing}},
  \bibnamefont{et~al.}, \bibinfo{journal}{Nat. Nanotech.}
  (\bibinfo{year}{2016}).

\bibitem[{\citenamefont{Sch{\"a}fer and
  Wegener}(2013)}]{schafer2013semiconductor}
\bibinfo{author}{\bibfnamefont{W.}~\bibnamefont{Sch{\"a}fer}} \bibnamefont{and}
  \bibinfo{author}{\bibfnamefont{M.}~\bibnamefont{Wegener}},
  \emph{\bibinfo{title}{Semiconductor optics and transport phenomena}}
  (\bibinfo{publisher}{Springer Science \& Business Media},
  \bibinfo{year}{2013}).

\end{thebibliography}


\appendix
\section{Tight-binding form of Hamiltonian}

Here we discuss the specific form of the Hamiltonian describing the class of perovsite materials.  Taking the example of the mixed halide perovskite CH$_3$NH$_3$PbI$_3$, its cubic form has direct band gap at $R$ point [$2\pi/a, 2\pi/a, 2\pi/a$], where $a$ is the lattice constant. The near-gap conduction band states are composed of the $p$ orbitals of Pb while the valence bands states are derived from the Pb $s$ orbital and I $p$ orbitals. The energy states near Fermi level can be described by a $8 \times 8$ tight-binding Hamiltonian including only $s$ and $p$ orbitals of Pb occuping the cubic lattice sites. We consider four types of inter- and intra-orbital hopping parameters, such as $t_{ss}^\sigma$, $t_{pp}^\sigma$, $t_{pp}^\pi$ and $t_{sp}^\sigma$. The effect of I $p$ orbitals is implicitly taken into account through tuning the magnitude of hopping parameters. The spin-orbit coupling splits degenerate conduction band states ($L=1$) into lower $J=1/2$ and upper $J=3/2$ bands, leading to a $J = 1/2$ conduction band and $S = 1/2$ valence band. In this study, we focus on the optical transition between valence and conduction bands so that we can truncate the $8 \times 8$ Hamiltonian to a $4 \times 4$ minimal continuum model to describe the near-gap optical transition. With the basis $\{ |S,\uparrow\rangle, |S,\downarrow\rangle,|J=1/2,j_z=+1/2\rangle,|J=1/2,j_z=-1/2\rangle\}$, the effective continuum Hamiltonian near the $R$ point up to second order in $k$  is given by

\begin{widetext}
\begin{eqnarray}
H = \left(
      \begin{array}{cccc}
        t_{ss}^\sigma k^2-\Delta/2 & 0 & i \xi k_z - \frac{1}{\sqrt{3}}\gamma_{sp}^z\left(k_+k_--4\right) & i \xi k_-  \\
        0 & t_{ss}^\sigma k^2+\Delta/2 & i \xi k_+ & -i \xi k_z + \frac{1}{\sqrt{3}}\gamma_{sp}^z\left(k_+k_--4\right) \\
        -i \xi k_z - \frac{1}{\sqrt{3}}\gamma_{sp}^z \left(k_+k_--4\right) & -i \xi k_- & \epsilon_0 + t_c k^2 + \Delta/6 & i\frac{4}{3}\gamma_{pp}^z k_- \\
        -i \xi k_+ & i \xi k_z+\frac{1}{\sqrt{3}}\gamma_{sp}^z\left(k_+k_--4\right) & -i\frac{4}{3}\gamma_{pp}^z k_+ & \epsilon_0 + t_c k^2 - \Delta/6 \\
      \end{array}
    \right)\nonumber\label{eq:4bandH}
\end{eqnarray}
\end{widetext}
where $k_{\pm}=k_x \pm i k_y$, $\xi=2t_{sp}^\sigma/\sqrt{3}$, $\gamma_2=4\gamma_{pp}^z/3$, $\epsilon_0=\left(\epsilon_p-\epsilon_s\right) -2\left(t_{pp}^\sigma+2t_{pp}^\pi\right) - \lambda + 6t_{ss}^\sigma$, $t_c=\left(2 t_{pp}^\pi + t_{pp}^\sigma\right)/3$.  $\Delta$ is the exchange interaction and we assume the magnetization is along the $z$-direction.  $\lambda$ is the spin-orbit coupling parameter responsible for splitting the $J=1/2$ and $J=3/2$ bands.  $\gamma_1$ and $\gamma_2$ parameterize the spin-dependent hopping terms induced by inversion symmetry breaking along $z$ direction.  The $t_v$ parameter of the Hamiltonian given in the main text (Eqs. \ref{eq:Hc}-\ref{eq:Hvc}) are related to tight-binding parameters as: $t_v=t_{ss}^\sigma,~\alpha_v=\frac{4}{3}\gamma_{pp}^z,~\gamma_1=\gamma_{sp}^z/\sqrt{3}$.

When the band gap energy is larger than all other energy scales, Eq. \ref{eq:4bandH} can be projected on to separate $2\times 2$ dimensional Hamiltonians for conduction and valence band, as given in Eqs. \ref{eq:Hc}-\ref{eq:Hvc} of the main text.  The parameters entering these projected Hamiltonians are related to the hopping parameters given here as: $\alpha_c = \gamma_2$, while $\alpha_v$ takes the more complex form:
\begin{eqnarray}
\alpha_v = \frac{1}{k} \left(\frac{\left(k\xi + \gamma_1\left(k^2-4\right)\right)^2}{tk^2+\epsilon_0-|k|\gamma_2} + \frac{\left(k\xi - \gamma_1\left(k^2-4\right)\right)^2}{tk^2+\epsilon_0+|k|\gamma_2}\right). \label{eq:alphav} \nonumber\\
\end{eqnarray}
In the limit of small $k$, Eq. \ref{eq:alphav} simplifies to:
\begin{eqnarray}
\alpha_v = \frac{16\gamma_1}{\epsilon_0^2}\left(2\gamma_1\gamma_2 - \xi\epsilon_0\right).
\end{eqnarray}

\begin{table}
\begin{tabular}{|c|c||c|c|}
  \hline
  Parameter & Default value [eV] & Parameter & Default value [eV]\\ \hline
  $t_{ss}^\sigma$ & -0.25   & $t_{pp}^\sigma$  & 0.9  \\
  $t_{sp}^\sigma$ & 0.4     &  $t_{pp}^\pi$ & 0.15  \\
  $\gamma_{pp}^z$      & -0.2 (0)    &  $\gamma_{sp}^z$    & -0.25 (0) \\
  $\epsilon_p$    & 4.0     & $\epsilon_s$ & -1.5 \\
  $ \epsilon_0$   & 1.5       &  $\lambda$     & 0.1 \\
  \hline
\end{tabular}
\caption{\label{tab:table-name}Default tight-binding parameter values.  $\gamma_{1,2}$ take on the values listed when Rashba spin-orbit coupling is included, and equal 0 for vanishing Rashba spin-orbit coupling.  The default parameters lead to an effective Rashba parameter for the conduction band $\alpha_c=0.1~{\rm eV}\cdot {\rm nm}$ and $B_{\rm R}=0.55~{\rm eV}$, and a band gap value of $E_g=1.5~{\rm eV}$ (for the case of vanishing $\Delta$).  The magnetic exchange field $\Delta$ and lifetime broadening $\epsilon$ are varied, and are specified in individual figure captions.}
\end{table}

\section{Derivation of steady state density matrix}

Here we review the derivation of the optical Bloch equations \cite{schafer2013semiconductor}.  In the eigenstate-basis of the ground state and considering only valence-conduction interband transitions, the Hamiltonian for the system including the optical excitation is:
\begin{eqnarray}
H({\bf k}) &=& \left(\begin{array}{cccc}
      E_{u}^v & 0 & 0 & 0 \\
      0 & E_{d}^v & 0 & 0 \\
      0 & 0 & E_{u}^c & 0 \\
      0 & 0 & 0 & E_{d}^c
    \end{array}\right) + \left(\begin{array}{cccc}
      0 & 0 & v_{uu} & v_{ud} \\
      0 & 0 & v_{du} & v_{dd} \\
      v_{uu}^* & v_{du}^* & 0 & 0 \\
      v_{ud}^* & v_{dd}^* & 0 & 0
    \end{array}\right)\nonumber
\end{eqnarray}
Here $E_{u,d}^v$ ($E_{u,d}^c$) is the ${\bf k}$-dependent valence (conduction) band energy for the $u,d$ state.  The $(u,d)$ label corresponds to the spin direction of the eigenstate, which is parallel ($u$) or anti-parallel ($d$) to the ${\bf k}$-dependent effective magnetic field.  $v_{jk}$ denotes the optical field-induced coupling between state $j$ of the valence band and state $k$ of the conduction band.

For the density matrix $\rho$, we denote the valence (conduction) band density matrix as $\rho^v$ ($\rho^c$), and $P$ to denote electron-hole density matrix elements.  The structure of $\rho$ is:

\begin{eqnarray}
\rho &=& \left(\begin{array}{cccc}
      \rho^v_{uu} & \rho^v_{ud} & P_{uu} & P_{ud} \\
      \rho^v_{du} & \rho^v_{dd} & P_{du} & P_{dd} \\
      P_{uu}^* & P_{du}^* & \rho^c_{uu} & \rho^c_{ud} \\
      P_{ud}^* & P_{dd}^* & \rho^c_{du} & \rho^c_{dd}
    \end{array}\right).
\end{eqnarray}
In our analysis, we switch from the conduction-valence representation of the density matrix to a electron-hole picture, where the electron density matrix is $\rho^e=\rho^c$, and the hole density matrix is $\rho^h = 1-\rho^v$.

The equation of motion for the density matrix $\rho$ is given by:
\begin{eqnarray}
\frac{\partial \rho}{\partial t} = \frac{1}{i\hbar}\left( \left[\rho,H\right] - i\epsilon\left(\rho-\rho_{eq}\right)\right)~. \label{eq:eom}
\end{eqnarray}
Equation \ref{eq:eom} leads to the semiconductor Bloch equations.  Writing these perturbatively in ${\bf E}$, and assuming that at $t=0$, $\rho^v_{uu}=\rho^v_{dd}=1$ while other elements of the density matrix are 0 leads to the following equation for $P_{jk}$:

\begin{eqnarray}
\left(\partial_t + \frac{i}{\hbar}\left(E_j^v-E_k^c\right)  + \frac{1}{\tau}\right)P_{jk}(t) = -iv_{jk}(t)~. \label{eq:p1}
\end{eqnarray}
We assume the time dependence of the excitation is given by $\cos\left(\omega t\right)=\left(\exp\left(i\omega t\right)+\exp\left(-i\omega t\right)\right)/2$.  Equation \ref{eq:p1} is solved by Fourier transform techniques, and yields the following expression for the dipole density matrix element:

\begin{eqnarray}
P_{jk}(t) &=& -\frac{v_{jk}}{2}\left(\frac{\exp\left(i\omega t\right)}{\hbar\omega + \left(E_k^c-E_j^v\right) - i\epsilon} ~+ \right. \nonumber \\ &&~~~~~~~~~\left. \frac{\exp\left(-i\omega t\right)}{-\hbar\omega + \left(E_k^c-E_j^v\right) - i\epsilon}\right),\end{eqnarray}
where $\tau$ is the carrier scattering time, and the dipole matrix element $v$ is given by:
\begin{eqnarray}
v_{jk} = i\frac{\langle j | {\bf v} \cdot {\bf E} | k\rangle }{ E_j^v - E_k^c}~, \label{:eq:vA}
\end{eqnarray}
where the velocity operator is ${\bf v} = \frac{\partial H }{ \partial {\bf k}}$.

Near a resonance condition $\hbar\omega \approx E_k^c-E_j^v$ for a pair of valence/conduction bands.  In this case, the term with denominator $\hbar\omega-\left(E_k^c-E_j^v\right)$ has the maximal contribution.  The equation of motion for hole-hole density matrix is, to lowest order in $v$:

\begin{eqnarray}
\frac{\partial}{\partial t} \rho_{jk}^h(t) &=& i\sum_{\ell\in c}\left(  v_{j\ell}P^*_{k\ell} - v^*_{k\ell}P_{j\ell}\right) \nonumber \\ &&- \left(i\left(E_j^v-E_k^v\right) -\frac{1}{\tau}\right) \rho_{jk}^h(t).\nonumber \label{eq:h1}
\end{eqnarray}
The sum $\ell$ is over conduction band states, while the indices $j$ and $k$ correspond to valence bands.  Letting $\partial \rho^h/\partial t=0$ and only including terms with denominators of the form $\hbar\omega - \left(E_k^c-E_j^v\right)$ yields the following expression for the steady state $\rho^h_{jk}$:

\begin{eqnarray}
\rho_{jk}^h &=&  \frac{1}{\epsilon+i\left(E_j^v-E_k^v\right)}\sum_{\ell\in c} \frac{i}{4}\left(\frac{v_{j\ell}v_{k\ell}^*}{\hbar\omega - \left(E_\ell^c-E_j^v\right) - i\epsilon}  \right. \nonumber  \\ && \left. ~~~ -\frac{v_{k\ell}^*v_{j\ell}}{\hbar\omega - \left(E_\ell^c-E_k^v\right) + i\epsilon}\right). \label{eq:rhoA}
\end{eqnarray}
Eq. \ref{eq:rhoA} is the general form for the hole density matrix under optical excitation.  The electron density matrix has a similar form.  The factor of $1/4$ is derived from expressing $\cos\left(\omega t\right)$ in terms of exponentials.

Here we give the explicit form for the density matrix in the limit of small Rashba spin-orbit coupling.  The valence and conduction band eigenstates are spinors along the $\bf z$-direction;
\begin{eqnarray}
\psi_u = \left(
             \begin{array}{c}
               1 \\
               0 \\
             \end{array}
           \right)~~~
\psi_d = \left(
             \begin{array}{c}
               0 \\
               1 \\
             \end{array}
           \right)
\end{eqnarray}
The velocity interband matrix elements are given as:
\begin{eqnarray}
v_{uu,dd}&=& \frac{\mp i \xi\sin\left(\beta\right)}{E_{uu,dd}} \nonumber\\
v_{ud,du}&=& \frac{-\xi\left(1\pm\cos\left(\beta\right)\right)}{E_{ud,du}}~,
\end{eqnarray}
where $E_{\sigma\sigma'}=\delta_{\sigma\sigma'}+tk^2$ and $\delta_{\sigma\sigma'}$ is the difference in energy between the $\sigma$ valence band edge and the $\sigma'$ conduction band edge.

The explicit form for Eq. \ref{eq:rhoA} is then give below:

\begin{widetext}
\begin{eqnarray}
\rho_{uu(dd)}^h &=& \frac{i\xi^2\left(1\pm\cos\beta\right)}{4\epsilon}\Bigg(\frac{1 - \cos\beta}{E_{uu(dd)}^2\left(\hbar\omega-E_{uu(du)} - i\epsilon\right)} + \frac{1+\cos\beta}{E_{ud(dd)}^2\left(\hbar\omega-E_{ud(dd)} - i\epsilon\right)} \nonumber \\
&&~~~~~~~~~~~~~~~~~~~~-\frac{1-\cos\beta}{E_{uu(du)}^2\left(\hbar\omega-E_{uu(du)} + i\epsilon\right)} - \frac{1+\cos\beta}{E_{ud(dd)}^2\left(\hbar\omega-E_{ud(dd)} + i\epsilon\right)}\Bigg), \label{eq:rhouu} \\
\rho_{ud}^h &=& \frac{i\xi^2\sin\beta}{4\left(i\epsilon+\Delta\right)}\Bigg(\frac{1-\cos\beta}{E_{du}E_{uu}\left(\hbar\omega-E_{uu} - i\epsilon\right)} +\frac{1+\cos\beta}{E_{dd}E_{ud}\left(\hbar\omega-E_{ud} - i\epsilon\right)} \nonumber \\ &&~~~~~~~~~~~~~~-\frac{1-\cos\beta}{E_{du}E_{uu}\left(\hbar\omega-E_{du} + i\epsilon\right)} - \frac{1+\cos\beta}{E_{dd}E_{ud}\left(\hbar\omega-E_{dd} + i\epsilon\right)} \Bigg), \label{eq:rhoud}
\end{eqnarray}
Integrating Eqs. \ref{eq:rhouu}-\ref{eq:rhoud} over ${\bf k}$ yields the final density matrix.  These $k$ integrals are of the general form:

\begin{eqnarray}
\int_0^\infty dk \frac{k}{\left(c_1+t k^2\right)\left(c_2+t k^2\right)\left(\omega^\pm - c_1-t k^2   \right)} =
\frac{1}{2 t \omega^\pm\left(c_1-c_2-\omega^\pm\right)} \left[\frac{\omega^\pm \ln \left(c_1/c_2\right)}{c_1-c_2} - \ln\left(\frac{\omega^\pm-c_1}{c_1}\right)\right],&&\nonumber \\
\end{eqnarray}
where $\omega^\pm=\omega \pm i \epsilon$, $t=t_{ss}^\sigma+t_c$.  The constants $c_{1,2}$ correspond to band splittings $\delta_{\sigma\sigma'}$.

We first present the resulting expressions for $n$ and $s_z$, which involve the diagonal elements of $\rho^h$.

\begin{eqnarray}
n &=& \left(\frac{\pi\xi^2 E^2}{4 t \epsilon \omega^2}\right)\left(\sin^2\beta\left(D_{uu}+D_{dd}\right) + \left(1+\cos\beta\right)^2D_{ud} + \left(1-\cos\beta\right)^2D_{du}\right), \label{eq:n}\\
s_z^h &=&\left(\frac{\pi\xi^2 E^2}{4 t \epsilon \omega^2}\right) \left(\sin^2\beta\left(D_{uu}-D_{dd}\right) + \left(1+\cos\beta\right)^2D_{ud} - \left(1-\cos\beta\right)^2D_{du}\right), \label{eq:sz}
\end{eqnarray}

$D_{\sigma\sigma'}$ is proportional to the joint density of states of the $\sigma$ valence band and the $\sigma'$ conduction band.  For the 2-d system considered here:
\begin{eqnarray}
D_{\sigma\sigma'} = \pi+2~{\rm Re} \left[\tan^{-1} \left(\frac{\hbar\omega - \delta_{\sigma\sigma'}}{\epsilon}\right)\right]
\end{eqnarray}
$\delta_{\sigma\sigma'}$ is the band gap splitting between valence band $\sigma$ and conduction band $\sigma'$, and $D$ varies between 0 and $2 \pi$.

The transverse spin density $s_{x,y}^h$ is determined by the off-diagonal element of the density matrix:

\begin{eqnarray}
s_x^h &=& \left(\frac{\pi\xi^2 E^2}{4 t \epsilon \omega^2}\right)\frac{\sin\beta}{1+\left(\Delta^v/\epsilon\right)^2}
\bigg[\left(\Delta^v/\epsilon\right) A + \omega B\bigg]\label{eq:sxfull}\\
s_y^h &=& \left(\frac{\pi\xi^2 E^2}{4 t \epsilon \omega^2}\right)\frac{\sin\beta}{1+\left(\Delta^v/\epsilon\right)^2}
\bigg[ A + \left(\Delta^v/\epsilon\right) \omega B\bigg] \label{eq:syfull}
\end{eqnarray}
The $A$ term has the same mathematical origin as the imaginary part of the dielectric function (e.g. the imaginary part of the retarded green's function), and is associated with absorption:
\begin{eqnarray}
A&=&\left(1-\cos\beta\right)\left(D_{uu}+D_{du}\right) + \left(1+\cos\beta\right)\left(D_{ud}+D_{dd}\right)
\end{eqnarray}
The $B$ term is peaked at photon energies which correspond to band edge transitions:
\begin{eqnarray}
B &=& \left(1+\cos\beta\right)\left(\frac{\left(\omega + \delta_{ud} - \delta_{dd}\right)\ln\left(\frac{\left(\delta_{dd}-\omega\right)^2+\epsilon^2}{\delta_{ud}^2}\right)}{\left(\omega + \delta_{ud} -\delta_{dd}\right)^2+\epsilon^2}
-\frac{\left(\omega + \delta_{dd} - \delta_{ud}\right)\ln\left(\frac{\left(\delta_{ud}-\omega\right)^2+\epsilon^2}{\delta_{dd}^2}\right)}{\left(\omega + \delta_{dd} -\delta_{ud}\right)^2+\epsilon^2}\right)+
 \nonumber \\
 &&~~
\left(1-\cos\beta\right)\left(\frac{\left(\omega + \delta_{uu} - \delta_{du}\right)\ln\left(\frac{\left(\delta_{du}-\omega\right)^2+\epsilon^2}{\delta_{uu}^2}\right)}{\left(\omega + \delta_{uu} -\delta_{du}\right)^2+\epsilon^2}
-\frac{\left(\omega + \delta_{du} - \delta_{uu}\right)\ln\left(\frac{\left(\delta_{uu}-\omega\right)^2+\epsilon^2}{\delta_{du}^2}\right)}{\left(\omega + \delta_{du} -\delta_{uu}\right)^2+\epsilon^2}\right)
\end{eqnarray}
\end{widetext}

In the limit where the excitation energy exceeds all band splittings $\hbar\omega\gg \delta_{\sigma\sigma'}$, $D_{\sigma\sigma'}=2\pi$ for all $\sigma,~\sigma'$, and the $B$ is negligible.  In this case, the hole spin density is related to the charge density via:
\begin{eqnarray}
{\bf s}^h = \frac{n}{2}\left(\left({\bf L}\cdot{\bf z}\right){\bf z}+ \frac{\left({\bf z}\times{\bf L}\times{\bf z}\right) + \left(\Delta^v/\epsilon\right) \left({\bf L\times\bf z}\right)}{1+\left(\Delta^v/\epsilon\right)^2} \right)
\end{eqnarray}
where $\bf z$ is the magnetization direction.  A similar analysis and result holds for electrons: the expressions given above are the same except for the replacement $\Delta^v\rightarrow \Delta^c$.  The forms of these expressions depend on the system dimensionality.  We've presented the 2-dimensional forms here, the 3-dimensional forms can be derived similarly.

\end{document}